**Investigation of boulder distribution in (1) Ceres and insight into its surface evolution**

Authors
**Mitsuha Noma [a], Naoyuki Hirata [b, *]**

* Corresponding author email address: hirata@tiger.kobe-u.ac.jp

**Authors' affiliation**
[a] Graduate School of Science, The University of Tokyo, Tokyo, Japan.
[b] Graduate School of Science, Kobe University, Kobe, Japan.

**Abstract**

The surface conditions of terrestrial bodies strongly reflect their geological evolutionary processes and vary among various terrestrial bodies. This diversity is attributed to variations in the timescales of boulder formation through processes such as impact cratering, rockfalls from crater walls, seismic motion, and boulder fragmentation caused by micrometeoroid impacts and thermal stress. In this study, we examined boulders on Ceres using high-resolution images with a resolution of approximately 5 m/px obtained during the Ceres Extended Mission 2 Orbit 7 of the Dawn mission. Almost all boulders were present around impact craters, even at a resolution of 5 m/px, thus indicating that the boulders on Ceres were created by impact cratering alone. The maximum boulder size on Ceres is approximately 200 m, even around large craters, which may indicate the upper size limit determined by the mechanical strength of the boulders, such as the tensile strength, the scale effect, and/or shattering strength. The slope of the size-frequency distribution of boulders on Ceres varied significantly across the range of boulder sizes, thus making it difficult to describe it using a single function of a power-law relationship; in particular, it changed at approximately 100 m, thus indicating that destructive or formation mechanisms may be different for large boulders > 100 m and for small boulders < 100 m. There may also be a subsurface structure that prevents the formation of small boulders, although this is difficult to argue conclusively. We estimated that the lifetime of boulders larger than 50 m was equivalent to or shorter than 100 Myr. This lifetime is consistent with a theoretical estimation assuming that micrometeoroid impacts are the primary destructive mechanism.

**Highlights**

- We counted boulders (> 15 m) to understand the surface evolution of Ceres.
- Almost all boulders on Ceres were formed by impact cratering.
- The largest boulders on Ceres (approximately 200 m) implied surface strength limits.
- The slope of the boulder size-frequency distribution changed at approximately 100 m.
- The lifetime of boulders on Ceres was equivalent to or shorter than 100 Myr.

## 1. Introduction

Ceres, with a diameter of 945 km, is one of the largest objects in the main asteroid belt. In 2015, the Dawn spacecraft arrived at Ceres and conducted various investigations (Russell and Raymond, 2012; McCord and Castillo-Rogez, 2018). Ceres is composed of a mixture of ice and rock (De Sanctis et al., 2015, 2018; Prettyman et al., 2016), and its interior is estimated to consist of a rocky core surrounded by an icy mantle (Park et al., 2016). Ceres has various surface features associated with impact cratering or endogenic activity. The surface of Ceres exhibits a heterogeneous crater distribution, an apparent absence of large craters, and a deficiency of surface water ice, despite the thermochemical models that have predicted that Ceres should have an icy crust with few impact craters (Buczkowski et al., 2016; Hiesinger et al., 2016; Toyokawa et al., 2022). Additionally, similar to other terrestrial bodies, numerous boulders are observed in and around impact craters on Ceres (Schröder et al., 2021).

Boulders on terrestrial bodies are associated with various surface processes, including physical properties, surface conditions, and regolith production, and are thought to reveal geological evolutionary processes. Boulders on terrestrial bodies other than the Earth were first resolved using Ranger photographs in the 1960s, which showed that boulders appeared around a small crater (Kuiper, 1965). Boulders are ubiquitous on terrestrial bodies, and their size and spatial distributions have been studied by many researchers; for example, the Moon (Shoemaker & Morris, 1970; Hörz et al. 1975 a, b; Cintala and McBride, 1995; Bart and Melosh, 2010; Watkins et al., 2019), Mars (Binder et al., 1977), Martian moons (Lee et al. 1986), Titan (Tomasko et al., 2005), asteroid (243) Ida (Lee et al., 1996; Sullivan et al., 1996), (433) Eros (Veverka et al., 2000, Thomas et al., 2001, Chapman et al. 2002; Dombard et al., 2010), (25143) Itokawa (Michikami et al., 2008), (162173) Ryugu (Michikami et al., 2019; Michikami and Hagermann, 2021), (21) Lutetia (Küppers et al., 2012), (4179) Toutatis (Jiang et al., 2015), (101955) Bennu (Dellagiustina et al., 2019; Uribe-Suárez et al., 2021), Dimorphos (Daly et al. 2023), (4) Vesta (Schröder et al., 2020), Ceres (Schröder et al., 2021), comet 67P/Churyumov-Gerasimenko (Pajola et al., 2015), 103P/Hartley 2 (Pajola et al., 2016), icy satellites (Hirata et al. 2022), and Enceladus (Martens et al., 2015, Pajola et al., 2021).

Most lunar and martian boulders are created by ejection from impact cratering, because they appear in or around impact craters (Shoemaker, 1965; Melosh, 1984; Krishna and Kumar 2016). Some boulders originate from the walls of impact craters or tectonic structures such as grabens due to rockfalls and the gravitational settling of the rock that occurs subsequent to seismic disturbances or mass wasting (Melosh, 1989; Miyamoto et al., 2007; Kumar et al. 2016; Ikeda et al., 2022; Grindrod et al., 2021). Boulders on Itokawa and Ryugu have been proposed to have resulted from catastrophic disruptions of the parent body (Saito et al., 2006, Fujiwara et al., 2006, Michikami et al., 2008; Michel et al., 2020). In addition to impact-related processes, periglacial activity on Mars (McEwen et al., 2007) and cryovolcanic eruptions and condensation around surface vents on Enceladus (Martens et al. 2015; Pajola et al., 2021) have been proposed. On the other hand, boulders mainly undergo fragmentation through two processes: (i) impact cratering and micrometeorite bombardment (Hörz and Cintala, 1997; Nakamura et al., 2008; Basilevsky et al., 2013; Ghent et al., 2014; Ballouz et al., 2020) and (ii) thermal stresses (Molaro and McKay, 2010; Viles et al., 2010; Molaro and Byrne, 2012; Delbo et al., 2014; Molaro et al., 2017; Hirata et al. 2022). Numerous small particles (micrometeorites) frequently impact planetary surfaces. The bombardment of such micrometeorites, which is an active process in the solar system, plays a major role in generating regolith and removal of boulders. For example, the typical lifetime of centimeter-to-meter-sized boulders on the Moon has been estimated to be $10^6$–$10^8$ years (Hörz and Cintala, 1997; Basilevsky et al., 2013, Ghent et al., 2014). Moreover, thermal stress is considered an important mechanism in the weathering and fragmentation of boulders. Boulders exposed to periodic diurnal and seasonal temperature variations develop sufficient thermal stress to induce thermal fatigue crack growth, which leads to regolith generation and boulder removal. Thermal fragmentation of a 10-cm rock on the Moon or asteroids in the main belt is predicted to occur at orders of magnitude faster than fragmentation due to micrometeoroid impact (Delbo et al., 2014). Thermal stress plays an important role in dust production and particle size reduction in the rings of Saturn over a short timescale (Hirata et al. 2022). Overall, the timescales of boulder formation and fragmentation depend on the material properties (such as tensile strength, porosity, and thermal conductivity) and environmental conditions of the surface, leading to significant variations in the density and distribution of boulders among terrestrial bodies.

Schröder et al. (2021) examined the size and spatial distribution of boulders on Ceres using low-altitude mapping orbit (LAMO) images obtained by the Dawn spacecraft, compared the boulder distributions of Ceres and Vesta, and concluded that all Ceres boulders were associated with craters and the typical lifetime of boulders on Ceres was shorter than that on Vesta. The LAMO images captured by the framing camera (FC) onboard the Dawn spacecraft (Russell et al., 2007) have a typical resolution of 35 m/px and cover almost the entire surface of Ceres (Roatsch et al., 2017). Schröder et al. (2021) detected boulders resolved to a size of at least three pixels (corresponding to 105 m on the surface of Ceres) in the LAMO images. On the other hand, the Dawn spacecraft obtained high-resolution images at a resolution of approximately 5 m/px during an extended mission called the Ceres Extended Mission 2 Orbit 7. This study aimed to examine boulders using the high-resolution images, reveal the distribution of small boulders with a size of less than 105 m, and compare our results to the size and spatial distribution of boulders on various terrestrial bodies to thereby provide further insights into the surface evolution of Ceres.

## 2. Methods

Two types of FC images were used for boulder counting (Fig. 1a, b): (i) images captured at an altitude of 385 km with a resolution of 35 m/px (i.e., the LAMO images) and (ii) images with a higher resolution obtained during the Ceres Extended Mission 2 Orbit 7. The former covered almost the entire surface of Ceres, while a portion of the polar regions (higher than 80° N or 80° S) was not clear. The latter were spread roughly within a small region from 60° S to 60° N and from 200° E to 260° E, and their imaging area was equivalent to 4.9% of the surface area of Ceres. The high-resolution images in the latter were unified into a single regional mosaic map with a resolution of 5 m and released by the German Aerospace Centre (DLR). To count boulders, we utilized the regional mosaic map. Hereafter, we refer to boulders identified from the former and the latter as the 35 m/px and 5 m/px-resolved boulders, respectively. The original images obtained during the extended mission vary in resolution and illumination and include areas that are not suitable for detecting boulders, such as shadows within craters, which would affect our counting. More importantly, the apparent boulder sizes are highly affected by the viewing geometry, such as emission angle (Li and Zhao 2023). Because the emission angles of the original images obtained during the extended mission were mostly below 20° (Fig. 2) and almost all of the boulders we identified appeared on a flat area (not on a crater wall), the shape of Ceres and viewing geometry hardly affected the boulder sizes in this study.

Boulder counting was performed using a tool called JMARS (Java Mission-planning and Analysis for Remote Sensing) (Christensen et al., 2009) that describes the positions (latitude and longitude) and sizes of boulders as three points, although this tool was developed for crater counting (Fig. 1c, d). Because most of the boulders we identified were partially lit, it was difficult to distinguish between their long and short axes (Fig. 3). Then, we placed the three points along the lit edge of each boulder. It should be noted that this method is adoptable for circular features, although most boulders are more or less irregular. The circles are then fitted to a roughly representative boulder. Therefore, the boulder sizes in this study indicate the mean diameter rather than the long axis of each boulder. We defined a boulder as an isolated positive-relief feature with a sharp edge.

Similar to Schröder et al. (2021), we define the minimum number of pixels for boulder identification as 3 pixels. Consequently, 105 m and 15 m were set as the minimum boulder diameters for detection in the 35 m/px and 5 m/px images, respectively. Nonetheless, it is generally said that in the case of an impact crater, the minimum number of pixels required for reliable identification is 10 pixels. In fact, boulders were clearly identifiable when they were resolved at greater than 10 pixels (Fig. 1h); however, they were obscure and their detection was ambiguous when we used 35 m/px images (Fig. 1g). In particular, boulders resolved at several pixels can be confused with another type of positive relief such as a hill. Based on the shadow, edge, and shading of each positive relief, we distinguished between hills and boulders; we defined that a boulder has an irregular shape and sharp edge, while a hill exhibits a smooth surface and its edge is obscure and extends toward its surroundings continuously and smoothly. It should be noted that boulders with sizes close to the minimum threshold should have been influenced by our interpretations.

The cumulative boulder size distributions in the solar system generally follow a power-law (Hartmann, 1969) or Weibull distribution (Schröder et al. 2020, 2021). In fact, the boulder size-frequency distribution of Ceres cannot be described by a single function of a power-law relationship as described below. Although the Weibull distribution would be more appropriate as a fitting function, we obtained the power-law index of the boulder size-frequency distribution, because it has become customary in the literature to characterize the cumulative size distribution of boulders/craters in terms of its power-law index. Therefore, we assume that the number of boulders larger than the diameter, *d*, can be described as:

$$N(>d) = N_{tot}\left(\frac{d}{d_{min}}\right)^{\alpha}, \quad (1)$$

where $N(>d)$ is the number of boulders larger than $d$, $d_{min}$ is the diameter of the smallest observed boulder (the minimum diameter threshold), $\alpha$ is the power-law index, and $N_{tot}$ represents the number of boulders larger than $d_{min}$. The power-law index α was computed using the maximum likelihood (ML) method. We calculated the sum of the logarithm of the ratio of the diameter of the $i$-th boulder, $d_i$ ($i = 1, \ldots, N$), to the minimum boulder diameter, $d_{min}$, using the following equation:

$$\alpha = -N(> d) \left( \sum_i^{N(>d)} \ln \left( \frac{d_i}{d_{min}} \right) \right)^{-1}. \quad (2)$$

This method is calculated directly from the boulder diameter, and thus, it is statistically valid, regardless of the type of data (cumulative, incremental, or binning) (Newman, 2005; Clauset et al., 2009).

## 3. Results

### 3.1. Boulder counting

The numbers of 35 m/px and 5 m/px-resolved boulders identified were 4,339 and 8,616, respectively (Fig. 4). The sizes and locations of these boulders are provided in the Supplementary Material. Schröder et al. (2021) identified 4,423 boulders using 35 m/px images of Ceres, and our results are similar to those of their work. All boulders were observed around crater rims (within two crater radii from the rim of the parent crater) or on crater floors (Fig. 1e, 5), while none were found around mountains or tectonic troughs unrelated to the crater (Fig. 1f). This indicates that impact cratering is the single source of boulders on Ceres, even in the 5 m/px images, although some of the boulders near crater walls may have been exposed by later seismic events. Tables 1 and 2 list craters with the 35 m/px and 5 m/px-resolved boulders. The tables show the size and position of each crater, the number of boulders larger than 3, 4, or 10 pixels, and the maximum boulder diameter. For example, the Jacheongbi crater (Table 1), which had the most boulders resolved at 35 m/px, had 487 boulders larger than 105 m, while the Occator crater (Table 2), which had the most boulders resolved at 5 m/px, had 1,836 boulders larger than 15 m (21% of the total boulders). The numbering of unnamed craters between Unnamed2 and Unnamed48 in Table 1 corresponds to those defined by Schröder et al. (2021), while that between Unnamed49 and Unnamed333 in Tables 2 and 4 (see Discussion) is newly defined. Note that Unnamed75 and 93 are removed in Table 2, because the two craters do not have boulders > 15 m.

Craters with both the 35 m/px and 5 m/px-resolved boulders were few; only the four craters, the Occator crater (Fig. 5a), Unnamed30 crater (Fig. 5b), Azacca crater (Fig. 5c), and Urvara crater (Fig. 5d), had both. Other than the four craters, the craters listed in Table 2 had no 35 m/px-resolved boulders. Among the boulders counted at 35 m/px resolution, the number of boulders counted at 5 m/px resolution was 9 of 66 in the Occator crater and 20 of 102 in the Unnamed30 crater. It is clear that our criteria overestimated the number of boulders in the 35 m/px images. In fact, most of the 35 m/px-resolved boulders were other surface features such as small hills or crater rims, when we observed them in the 5 m/px images (Fig. 6). Additionally, the spatial density of the 5 m/px-resolved boulders around the four craters was one-tenth that of the 35 m/px-resolved boulders in the diameter range of > 200 m (Fig. 7c, 7d). Although the 5 m/px images did not cover the entire surface of Ceres, it is possible that most of the 35 m/px-resolved boulders were not boulders. We consider that the number of pixels for reliable identification is 10 pixels and at least 6 pixels. In other words, the 5 m/px-resolved boulders larger than 50 m in diameter were significantly reliable, while the 5 m/px-resolved boulders less than 50 m and 35 m/px-resolved boulders less than 350 m were not. We compared the 5 m/px-resolved boulders to boulders counted by Schröder et al. (2021). Schröder et al. (2021) identified 197 and 159 boulders in the Occator and Unnamed30 craters, respectively. The number of boulders common to our study and their boulders was 26 of 197 in the Occator and 63 of 159 in the Unnnamed30 craters. Among the common boulders, approximately half were similar in size to ours, while the rest were approximately twice as large. It should be noted that these boulders were resolved to a size of less than 5 pixels in 35m/px images, and none of the boulders were larger than 200 m (~6 pixels) in the small study area resolved by the high-resolution images. Therefore, our study cannot evaluate the reliability of the previous study.

### 3.2. Boulder size-frequency distribution

The cumulative size-frequency distribution of the boulders on Ceres (Fig. 7a, 7b) showed that its slope became shallower as the boulders became smaller, and the power-law index α of the 35 m/px-resolved boulders transitioned from -6.62 ($d$ > 175 m) to -3.95 ($d$ > 105 m), while that of the 5 m/px-resolved boulders from -1.84 ($d$ > 75 m) to -1.30 ($d$ > 15 m). The slope varied significantly across the range of boulder sizes, thus making it difficult to describe using a single function. Cintala and McBride (1995) also stated that the power-law index cannot describe the entire range from millimeters to meters for lunar boulders; as the size increases, the distribution becomes steeper, which may be due to ejection and reimpact with additional size sorting caused by weathering, erosion, and transport of materials. Fujiwara et al. (1989) stated that a single power-law index cannot describe the entire range in size for catastrophically disrupted fragments, the size distribution of the fragments is divided into two or three segments and exhibits a steeper slope for larger fragments, and the change in slope typically occurs at fragment sizes smaller than ~1/10 of the original target and is likely associated with the transition from plastic to elastic flow within the target. Fig. 7c presents the cumulative size-frequency distribution of the spatial density of the boulders. It appears that both functions converge to $\alpha = -1.87$ as the boulders became smaller ($d$ < ~100 m), the slope became steeper ($d$ > ~100 m) as the boulders became larger, and the slope changed at approximately 100 m. Fig. 7d shows the cumulative number density of boulders around some craters.

Fig. 8a presents the relationship between crater diameter and the power-law index $\alpha$ of the boulder size-frequency distribution. The absolute value of $\alpha$ for the 5 m/px-resolved boulder decreased as the crater diameter increased, with a correlation coefficient of 0.76. In contrast, for the 35 m/px-resolved boulders, a strong correlation was not observed (correlation coefficient of 0.18). Fig. 8b presents $\alpha$ for boulders on various terrestrial bodies. Our result for the 35 m/px-resolved boulders was close to that of Schröder et al. (2021), while for 5 m/px-resolved boulders differed significantly from those of other terrestrial bodies and Schröder et al. (2021).

### 3.3. Relationship between boulders and craters

Fig. 9a shows a comparison of the maximum boulder and crater diameters. Although the diameter of the largest boulder over Ceres was 460 m, based on our identification, most of the maximum boulder diameters were approximately 200 m, even around large craters. A positive correlation between the maximum boulder diameter and crater diameter was observed in smaller craters. An empirical relationship between the maximum boulder size and crater diameter has been proposed for the Moon, asteroids, and martian satellites (Moore 1971; Lee et al., 1996). A significant portion of the observed boulders was distributed above the empirical relationship. This size relationship for the Ceres crater, with a diameter between 300 m and 10 km, is several times larger than that reported by Lee et al. (1996).

Approximately 200 m appears to be the upper limit of the size of boulders in various solar system solid bodies. The maximum boulder diameters reported in previous studies by Lee et al. (1996), Moore (1971), Krishna and Kumar (2016), and Bart and Melosh (2007) were 150, 220, 63, and 410 m, respectively. Even for catastrophic disruptive objects such as Ryugu and Itokawa, the diameters of the largest boulders are 160 m (Otohime) and 50 m (Yoshinodai) (Michikami et al., 2008; Michikami et al., 2019). Almost no asteroids larger than a few 100 m spin faster than a rotation period of approximately 2.4 h (the so-called spin barrier), while smaller asteroids include fast rotators beyond the spin barrier (e.g. Harris 1996; Pravec and Harris 2000; Scheeres et al 2015). It is proposed that such fast rotators are in a strength-dominated regime and have a monolithic structure. Perhaps these sizes are the upper size limit determined by the strength of the target surface. It is well established that material strength depends on the sample size, which is known as the size effect or scale effect. As reviewed by Housen and Holsapple (1999), the tensile/compressive strength decreases with increasing sample size due to two mechanisms. (i) The larger sample includes weaker regions with more cracks, and the weakest spot determines the strength. (ii) The speed of crack growth depends on the rate at which a sample is loaded. Only small cracks have time to grow if loading occurs quickly, while larger cracks have time to coalesce if loading occurs slowly. Because the characteristic strain rates at the scaled ranges in a collision event vary inversely with the size of the impactor, the sample in a large-scale collision appears weaker than that in a small-scale event. Housen and Holsapple (1999) concluded that a kilometer-sized target requires approximately 500-fold less energy per mass for shattering than a meter-sized target, because of the two mechanisms. It is likely that the upper limit of the boulder size in nature is responsible for this scale effect.

## 4. Discussion

As all boulders on Ceres, even those resolved at 5 m/px, were observed around craters and absent in non-crater regions, the generation mechanism for the observed boulders on Ceres was thought to be solely impact cratering. This is further supported by the positive correlation observed between the crater diameter and the cumulative volume of boulders (Fig. 9b), where the total volume of boulders ($V_{boulder}$) is defined by:

$$V_{boulder} = \frac{\pi}{6} \sum_{j}^{M} {d_j}^3 \ , \qquad (3)$$

where $d_j$ (j=1,…,M) is the diameter of the boulders (> 3 px) associated with the crater, and the volume of each boulder is assumed to be equal to that of a sphere with a diameter of $d_j$. The minimum $d_j$ values were 105 m and 15 m for the 35m/px and 5m/px-resolved boulders, respectively. The correlation coefficient between crater diameter and the total volume of boulders was 0.84, thus supporting that these boulders originated from ejecta. However, lunar boulder masses approach 5% of the total excavation from source craters (Cintala et al. 1982), whereas the boulder masses on Ceres in our measurements were up to 0.01% of the total excavation from source craters, thus indicating that the volume fraction of boulders to source craters for Ceres is much smaller than that for the Moon. If the ratio of the total volume of boulders to the volume of their parent crater is independent of the crater size, we should take a relation of $V_{boulder} \propto D^3$, where $D$ is the crater diameter. However, we obtained a power-law index of 1.91 ($V_{boulder} \propto D^{1.91}$). This can be explained by the hypothesis that fewer smaller boulders are present on Ceres, as described below.

Reliable results, not affected by the minimum number of pixels for identification, were as follows; (i) the ratio of the maximum boulder size to the crater size for Ceres was several times larger than that for the Moon, asteroids, and martian satellites, and (ii) the slope of the boulder size-frequency distribution changed at approximately 100 m and became shallower as the boulders became smaller. The power-law index was -3.2 for boulders on Phobos with diameters exceeding several meters around the Stickney crater (Thomas et al. 2000), -3.2 for boulders on Eros in a diameter range of 15–80 m (Thomas et al. 2001), -2.5 to -2.7 for boulders on Eros within the Shoemaker crater (Thomas et al. 2001), -3.1 for boulders on Itokawa in a diameter range of 7–35 m (Michikami et al., 2008), -2.65 for boulders on Ryugu with diameters exceeding 5 m (Michikami et al., 2019), -3.6 for boulders on Churyumov-Gerasimenko with diameters exceeding 7 m (Pajola et al., 2015), and -2.2 for boulders on Churyumov-Gerasimenko within the neck region (the area between the main lobe and the small lobe) (Pajola et al., 2015). Compared to boulders on these objects, -1.87 for boulders on Ceres in the diameter range of 15–100 m was significantly shallow. Although this shallow distribution can be explained by either the paucity of smaller boulders or the paucity of larger boulders, we consider the former to be more likely, because a deficiency of large boulders is unlikely. Therefore, our results suggest that Ceres contains larger boulders and fewer small boulders than the Moon or other objects. The destructive mechanisms may be very different for large boulders > 100 m and for small boulders < 100 m, and smaller boulders were preferentially eliminated. Otherwise, a mechanism that prevents the formation of small boulders, possibly associated with subsurface structures, may be responsible for the paucity of small boulders. However, we are unable to provide a plausible explanation for these findings.

Boulders whose sizes are several times the thermal skin depth are generally the most susceptible to breakdown due to thermal stress (Molaro et al., 2017). The skin depth is a measure of the depth at which the amplitude of the periodic temperature variation is reduced to 1/e times its value at the surface. The theoretical skin depth in a semi-infinite half-space is expressed as follows (Turcotte and Schubert, 2002):

$$d_w = \sqrt{\frac{2k}{\rho C \omega}}, \quad (3)$$

where $k$ is the thermal conductivity, $C$ is the specific heat capacity, $\rho$ is the density of the target, and $\omega$ is the frequency of temperature change at the surface ($\omega = 2\pi/T_p$, where $T_p$ is the period of periodic temperature change). The axial tilt of Ceres is small, but the orbital eccentricity is large, which causes a temperature variation of 15 K during one orbital cycle (Formisano et al., 2016). Using the parameters $k = 0.02$ W m$^{-1}$ K$^{-1}$, $C = 6.39$ J kg$^{-1}$ K$^{-1}$, and $\rho = 2080$ kg m$^{-3}$ (Formisano et al. 2016), we can obtain $d_w = 0.83$ m. Therefore, thermal stress is unlikely to be a differential destructive mechanism with a diameter of 100 m.

We estimated the lifetimes of boulders on Ceres based on the ages of the parent craters with boulders. As the boulders were created by cratering, their ages should correspond to those of the parent craters. Specifically, we assumed that the majority of boulders were exposed when the parent crater formed and that the typical lifetime of boulders should be shorter than the ages of craters without boulders and longer than the ages of craters with boulders. It should be noted that some boulders may have been exposed by seismic events, mass movements, or various surface processes and that the ages of such boulders do not match the age of the parent crater. We utilized the crater ages determined by previous studies as summarized in Table 3 (Hughson et al. 2018; Krohn et al. 2018; Neesemann et al. 2019; Pasckert et al. 2018; Ruesch et al. 2018; Schmedemann et al. 2016; Schulzeck et al. 2018; Scully et al. 2018; Toyokawa et al. 2022; Williams et al. 2018). In these previous studies, the lunar-derived model (LDM) and/or asteroid-derived model (ADM) were used to relate crater number density to absolute age. According to literatures based on the ADM, the ages of the four craters are 46 Myr (Azacca), 13 Myr (Occator), 50 Myr (Unnamed30), and 110 Myr (Urvara). According to literatures based on the LDM, they are 76 Myr (Azacca), 42 Myr (Occator), 69 Myr (Unnamed30), and 242 Myr (Urvara). Table 4 lists craters without 5m/px-resolved boulders, such as the Nepen crater. Because the ages of these boulder-free craters have not been obtained in the previous studies, we counted small craters on the crater floors using 5 m/px images. Fig. 10 shows the crater density (> 200 m in diameter) on the floor of each crater and the boulder density (>50 m in diameter) around the crater. This indicates that the ages of the boulder-free craters are not significantly different from those of the Azacca and Unnamed30 craters. Therefore, it is reasonable that if we use the crater age based on the ADM, the lifetime of Ceres boulders (> 50m) would be equivalent to or shorter than approximately 100 Myr. If we use the LDM, it would be equivalent to or shorter than approximately 200 Myr. Our results are consistent with those of Schröder et al. (2021), which determined the lifetime of 100 m boulders on Ceres to be 150 Myr.

Before the arrival of the Dawn spacecraft, Basilevsky et al. (2015) theoretically estimated that the lifetime of meter-sized boulders on Ceres, assuming micrometeoroid impacts to be a primary destructive mechanism, would be approximately 10–70 Myr and larger boulders would be destroyed much more rapidly. Our result is slightly longer than but roughly consistent with this prediction. Therefore, it is reasonable to assume that micrometeoroid impacts are the primary destructive mechanism for boulders on Ceres. Schröder et al. (2021) proposed that boulders on Ceres are mechanically weak because they are composed of water ice. Our boulder counting itself does not determine if boulders on Ceres are weaker or stronger than those on other bodies; however, our estimation of the lifetime of Ceres boulders corresponds to one extrapolated from lunar boulders, which indicate that the strength of Ceres boulders is not significantly different from that of lunar boulders. Our results agree with those of an impact experimental study by Arakawa et al. (2022) that was published after Schröder et al. (2021). Arakawa et al. (2022) demonstrated that the shattering strength of icy clay samples with water contents of up to 35 wt% is not different from that of basalt, although the static tensile strength of basalt is approximately 8 times larger than that of the icy clay samples. Therefore, if the boulders on Ceres are of impact origin, it is reasonable that the shattering strength of the boulders on Ceres is not significantly different from that of lunar boulders.

## 5. Conclusion

We investigated the size and spatial distribution of boulders on Ceres using the 5m/px images obtained by the Dawn spacecraft to provide insight into the evolutionary history of Ceres. Consequently, we identified 8,616 boulders larger than 15 m in diameter, although our identification of boulders resolved at less than 10 pixels (corresponding to 50 m) may be unreliable. Almost all boulders were present around impact craters, which is further supported by a positive correlation between the total volume of boulders and the size of their parent craters. The maximum boulder size on Ceres was approximately 200 m (even around large craters), indicating the upper size limit determined by the mechanical strength of the target surface, such as the tensile strength, scale effect, and/or shattering strength. Although the maximum boulder sizes around craters on Ceres were much larger than those around craters on other terrestrial objects, the slope of the boulder size-frequency distribution on Ceres in the diameter range below 100 m was shallower than that on other terrestrial bodies, thus implying a paucity of small boulders below 100 m on Ceres. Possible explanations for the paucity of small boulders include the existence of a process that preferentially eliminates the smaller boulders or a mechanism that prevents the formation of the small boulders; however, we were unable to constrain them. We estimated that the lifetime of boulders larger than 50 m was equivalent to or shorter than 100 Myr. This lifetime is consistent with the theoretical estimation of Basilevsky et al. (2015), which assumed micrometeoroid impacts to be the primary destructive mechanisms.

**Acknowledgments**

We appreciate Akiko M. Nakamura who provided helpful comments on the upper limit of boulder size. We thank the two anonymous reviewers for their helpful comments that significantly improved our study. We used the regional mosaic maps of Ceres released via DLR (http://dawngis.dlr.de/atlas_dir/schema/Ceres_Schema_Mosaic.html) and JMARS to conduct the analyses. This work was partly supported by JSPS Grants-in-Aid for Scientific Research (Nos. 20K14538 and 20H04614) and the Hyogo Science and Technology Association.

**Supplementary material**

We uploaded csv files exported from JMARS to present a catalogue of the 35m/px-resolved bouldersand 5m/px-resolved boulders (

**Figures**

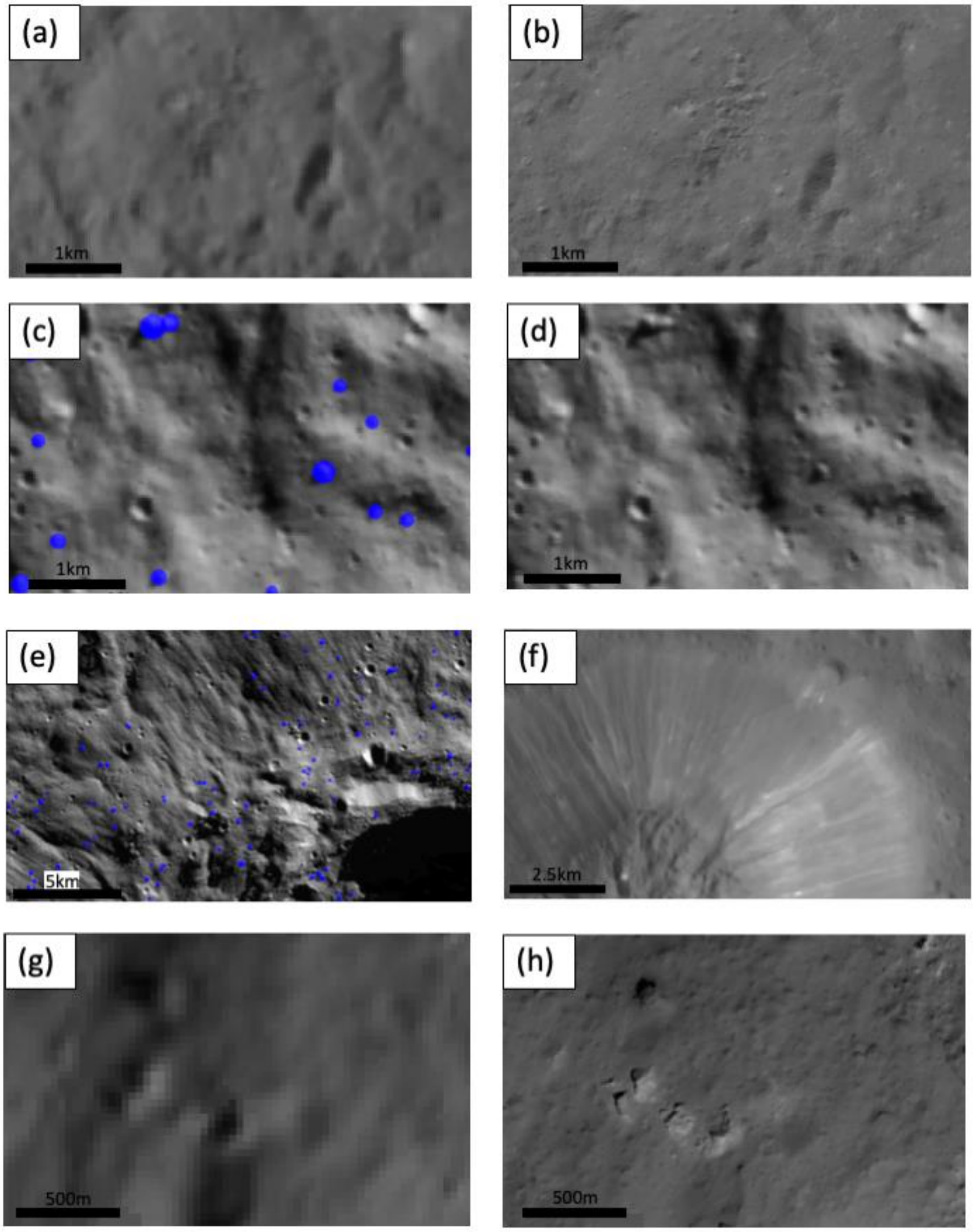

**Figure 1.** (a, b) Comparison between images at a resolution of (a) 35 m/px and (b) 5 m/px, showing the floor of the Occator crater. (c, d) Example of our boulder counting, where blue circles in (c) outline boulders that we have identified from the original image presented in (d). Both images have a resolution of 35 m/px. (e) Boulders are highlighted by blue circles around the Jacheongbi crater. (f) A region around the Ahuna Mons exhibits no boulders. (g, h) Difference in visibility of boulders at a resolution of (g) 35 m/px and

(h) 5 m/px. The center latitude and longitude of each location is (a, b) 17.8° N, 242.6° E, (c, d) 66.9° S, 0.81° E, (e) 67.2° S, 358.2° E, (f) 10.1° S, 317.0° E, and (g, h) 19.48° N, 244.4° E.

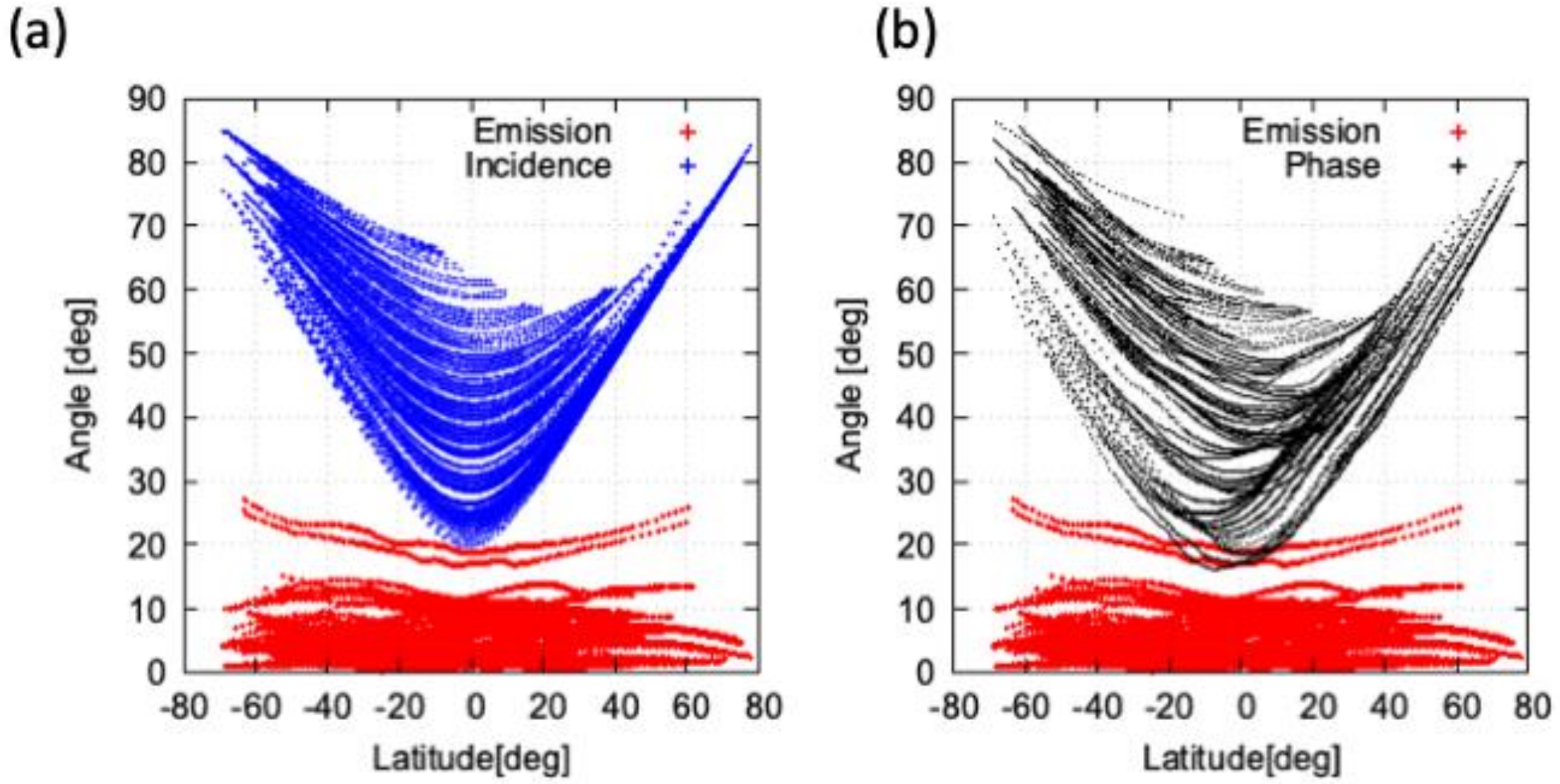

**Figure 2.** The emission, incidence, and phase angels of the high-resolution images. These data are from Nathues et al. (2016). The horizontal axis is the latitude of each image, and the vertical axis is the angles.

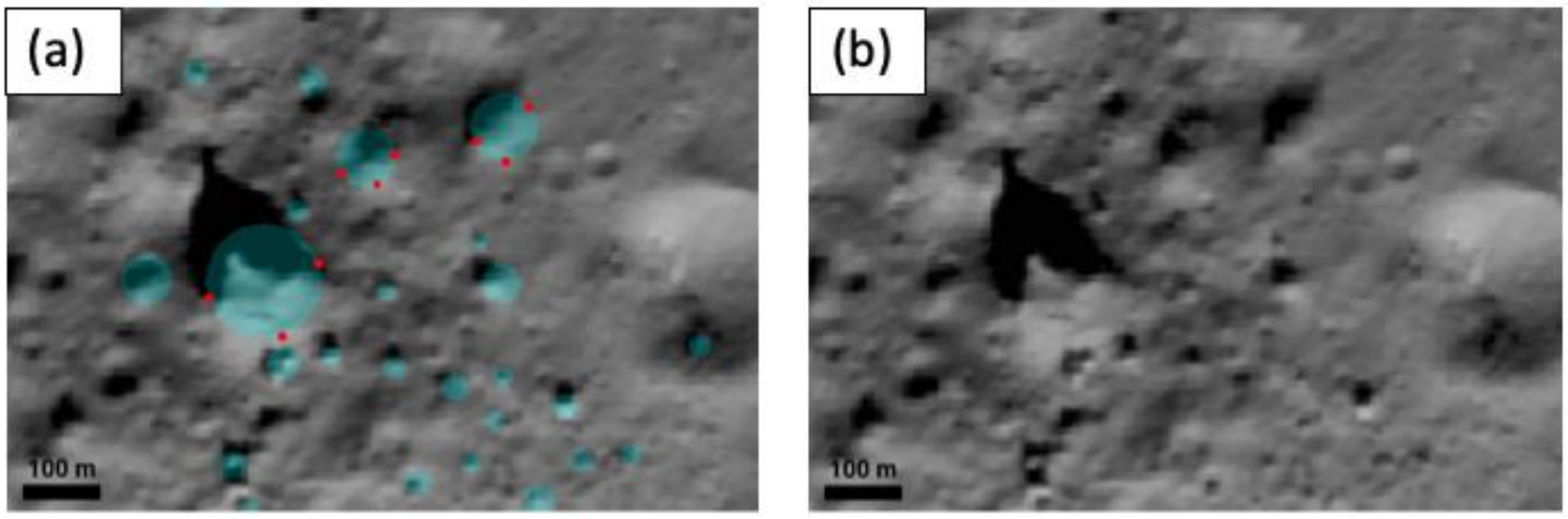

**Figure 3.** Examples of selection of the three points (red dots). The image has a resolution of 5m/px. The center latitude and longitude is 50.1°N, 239.8°E.

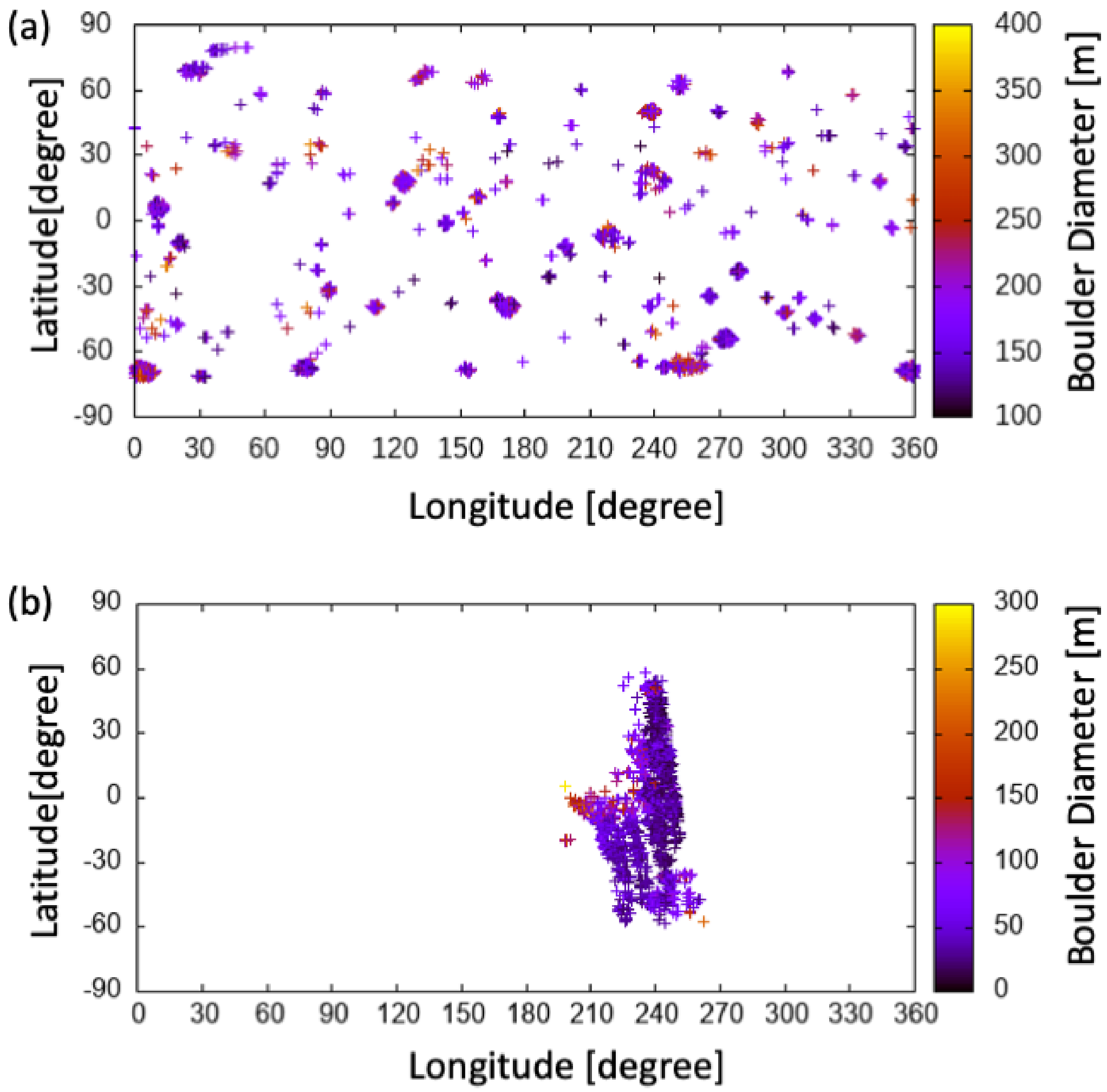

**Figure 4.** Global distribution of (a) the 35 m/px and (b) 5 m/px-resolved boulders.

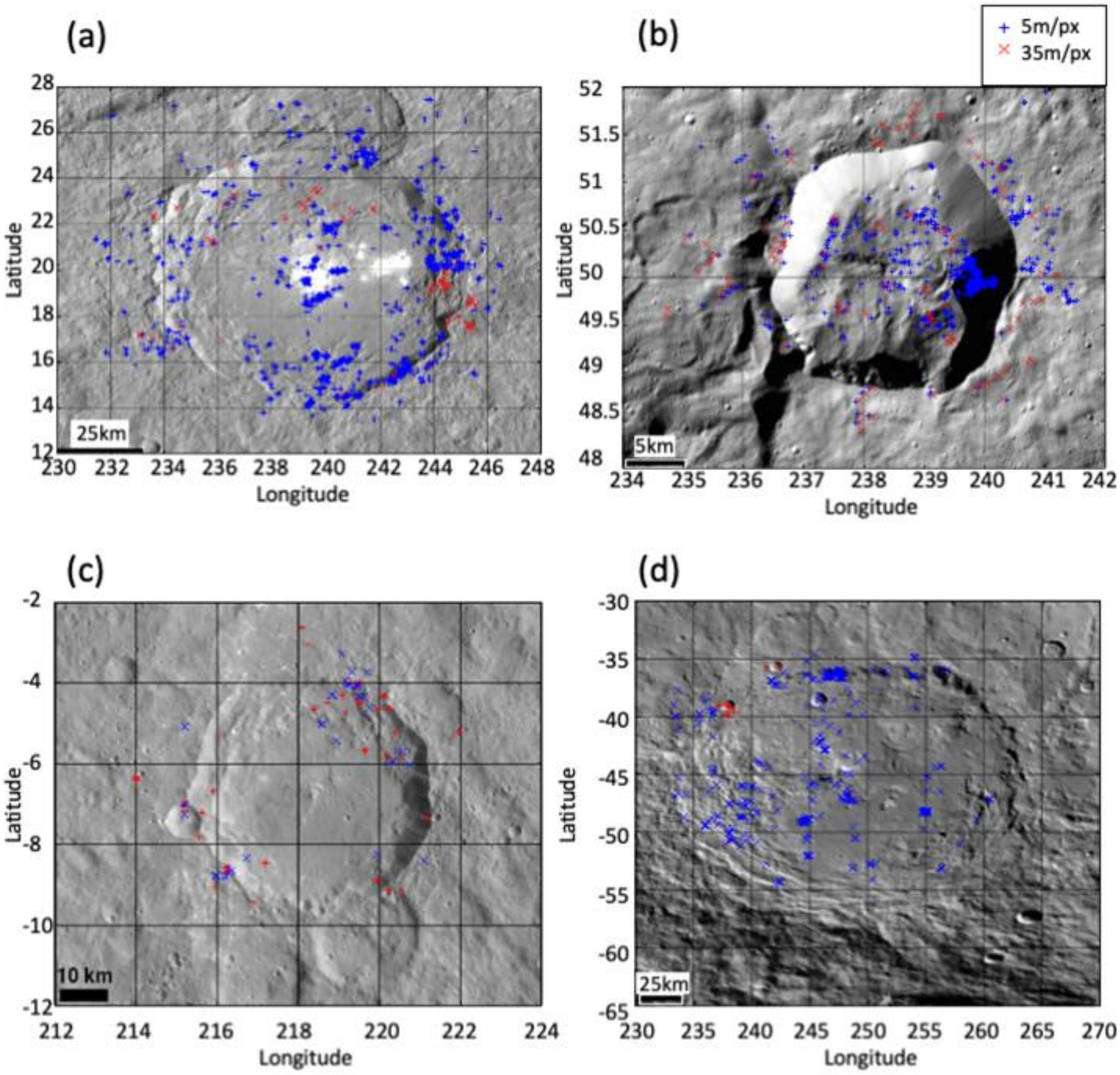

**Figure 5.** Location of boulders around (a) the Occator crater, (b) Unnamed30 crater, (c) Azacca crater, and (d) Urvara crater. Red and blue crosses represent the 35 m/px and 5 m/px-resolved boulders, respectively.

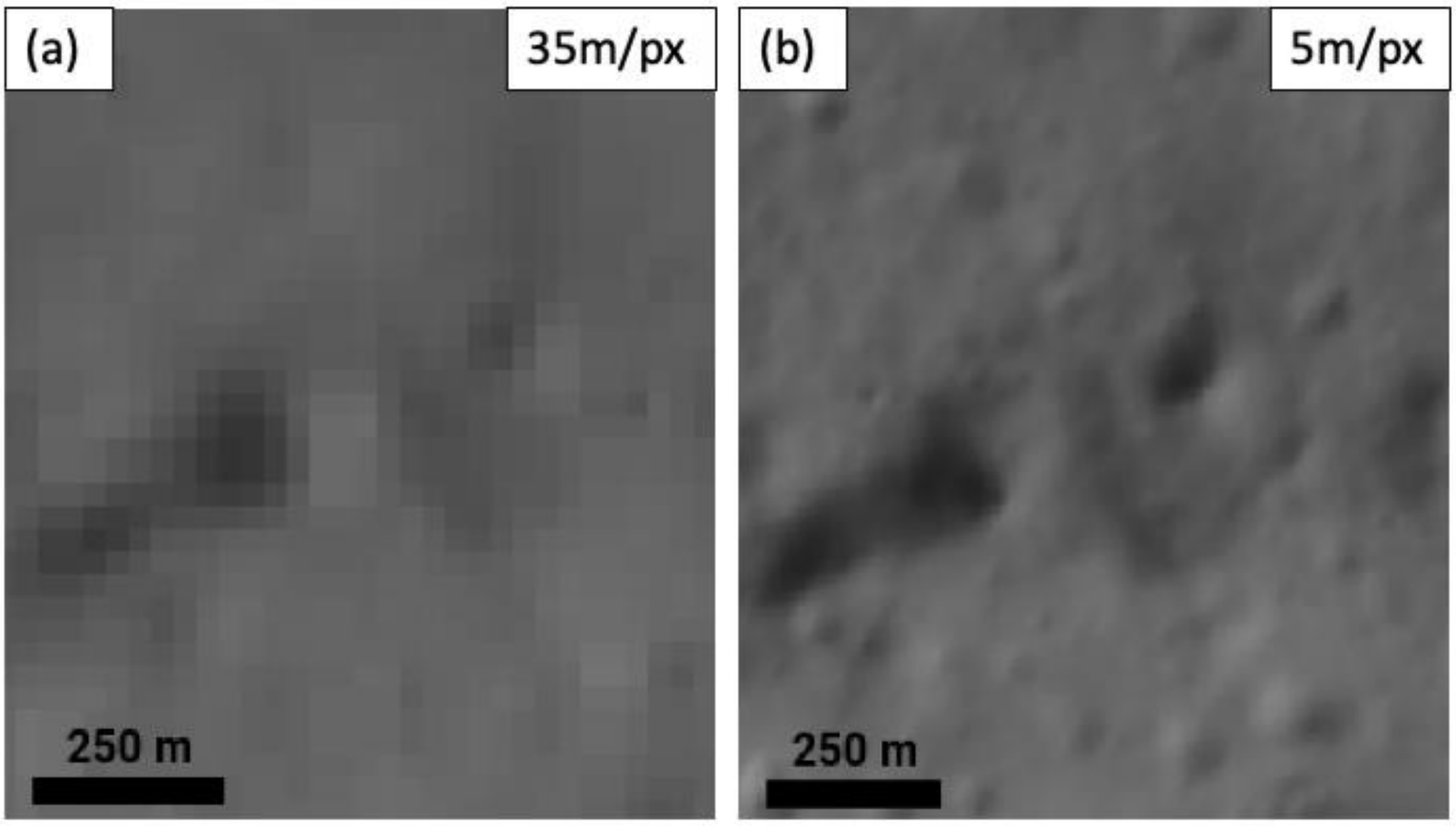

**Figure 6.** An example of a boulder that we regarded as boulder in the 35 m/px image (a) but not in a 5 m/px image (b). Small isolated positive relief features are small hills and not boulders at a resolution of 5 m/px. At a resolution of 35 m/px, the small positive features have 7 or 8 pixels. The small hills may be crater rims. The center latitude and longitude is 24.59$^{\circ}$ N, 244.4$^{\circ}$ E.

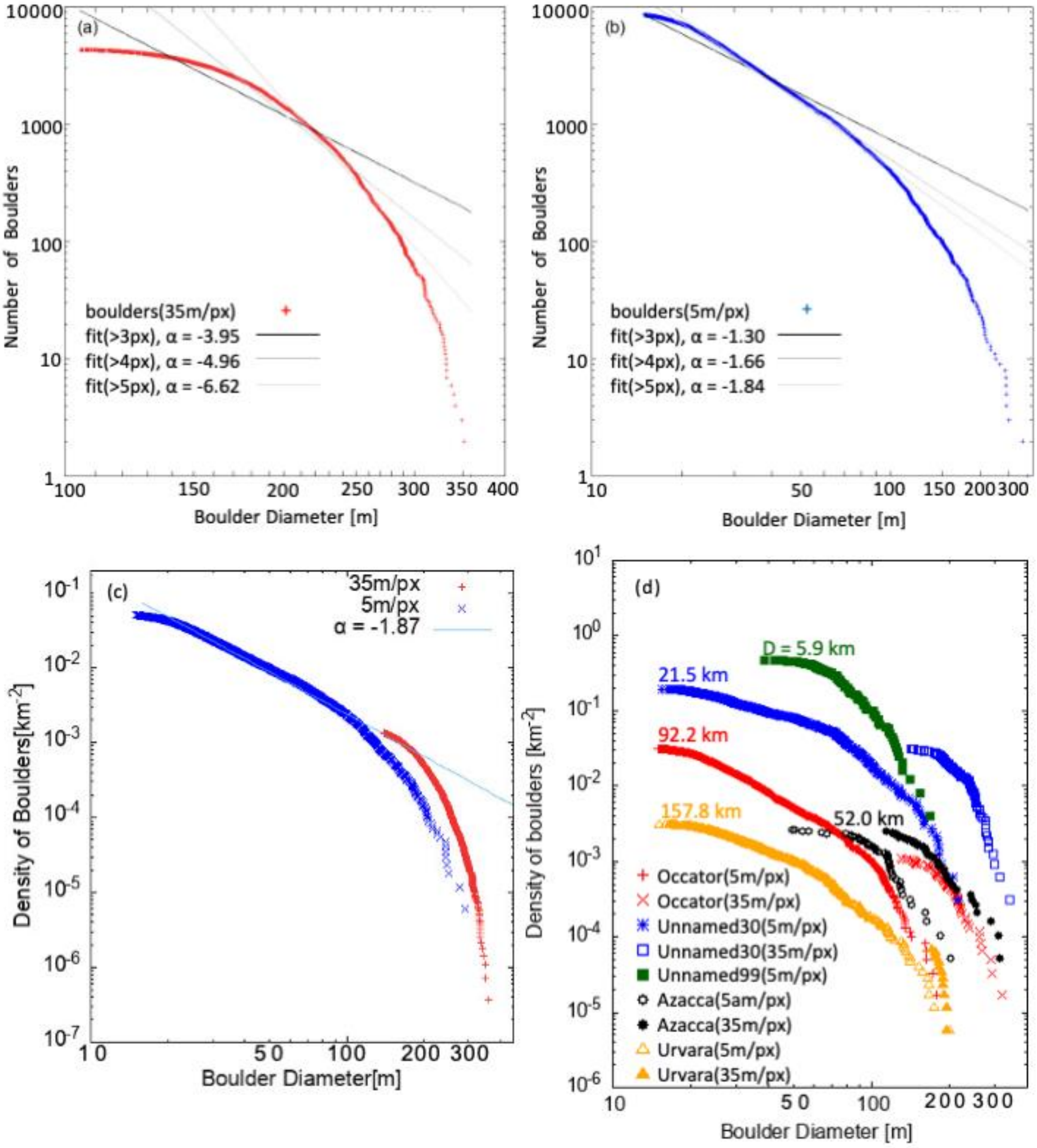

**Figure 7.** (a, b) Cumulative number distribution of (a) the 35 m/px and (b) 5m/px-resolved boulders on Ceres as a function of the boulder diameter. (c) The cumulative size-frequency distribution of the spatial density of the boulders. Here, we assume the areas to have a surface area of 473 km-radius sphere for the 35 m/px-resolved boulders and 4.9% of the surface area for the 5 m/px-resolved boulders. (d) Boulder size-frequency

distribution around representative craters. The adjacent numbers represent the diameter of the crater.

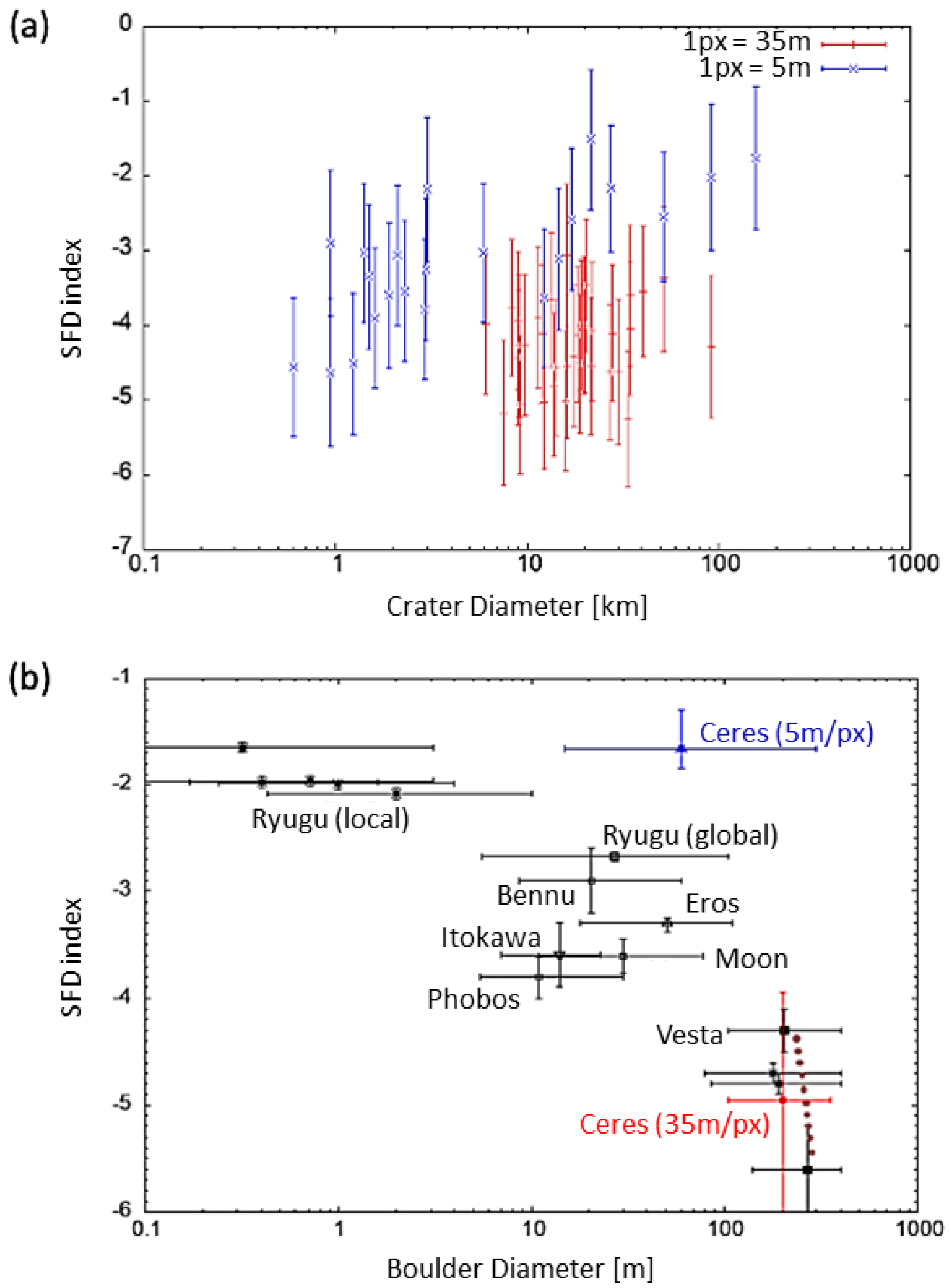

**Figure 8.** (a) Relationship between crater diameter and slope (α, index) of the size-frequency distribution of boulders. The error bars represent one standard deviation (±1σ). (b) Relationship between boulder diameter and slope of the boulder size-frequency distribution for various terrestrial bodies. This panel is adopted from Schröder et al. (2020). Red and blue crosses indicate our measurements.

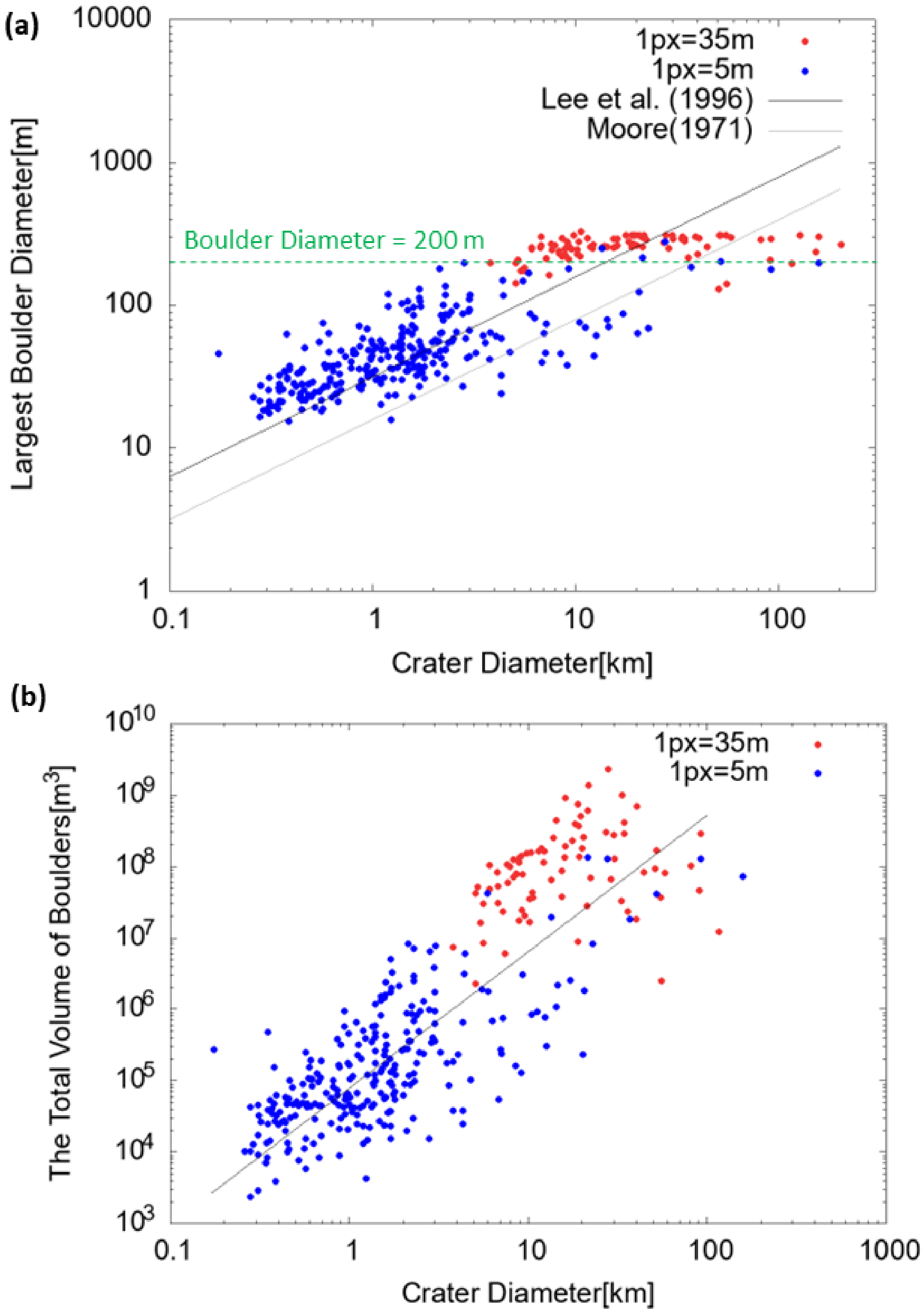

**Figure 9.** (a) Relationship between crater diameter and maximum boulder diameter. The black and gray lines show the empirical relationship proposed by Lee et al. (1996) and Moore (1971). (b) Relationship between crater diameter and total volume of boulders. The black line shows a power-law index of 1.91.

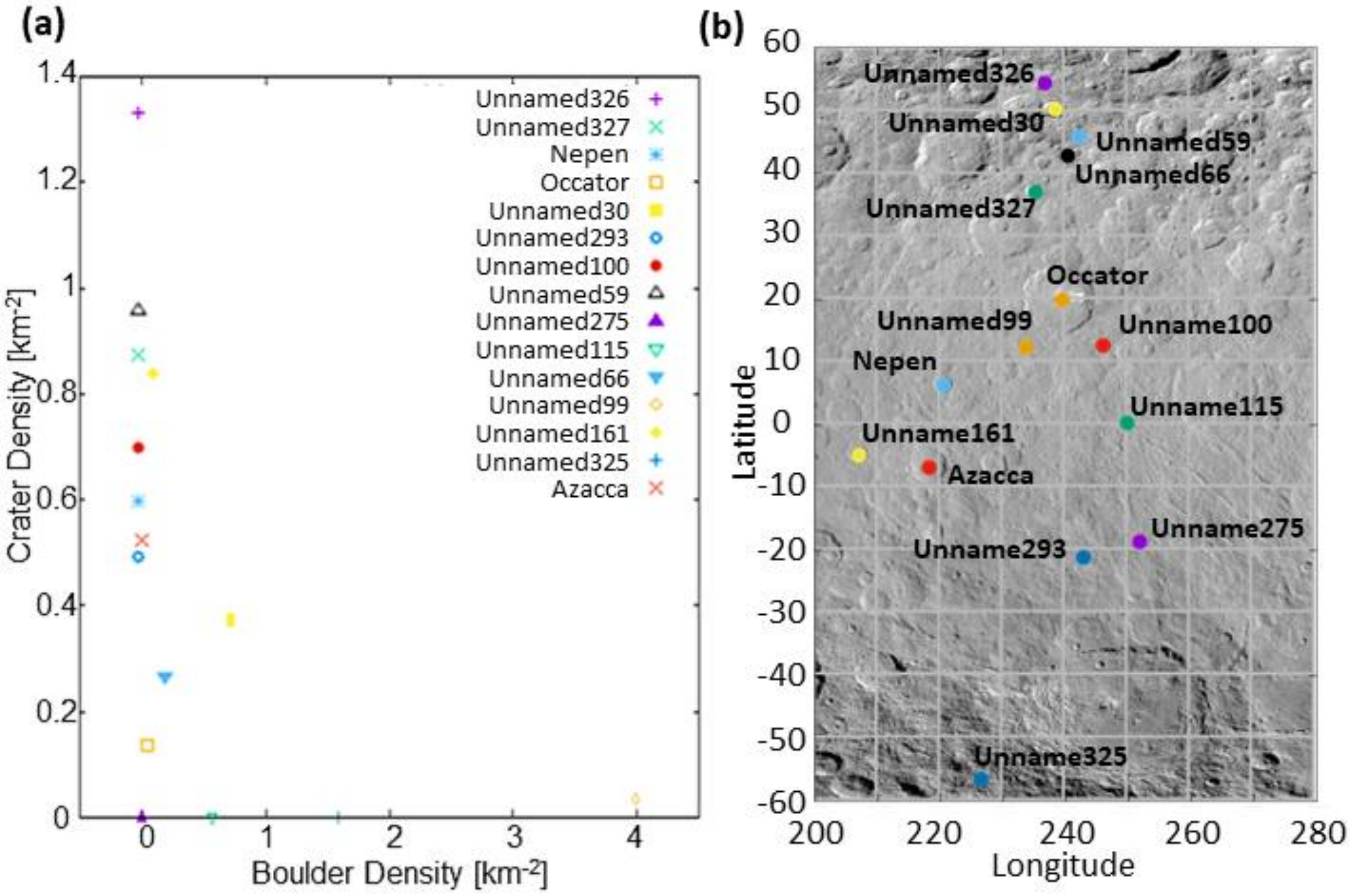

**Figure 10.** (a) Relationship between boulder number density and small-crater number density of each crater. The crater number density is defined as the number of craters larger than 200 m in diameter located within the crater floor. The boulder number density is defined as the number of boulders larger than 50 m in diameter located within the three crater radii from the center of their parent crater. (b) The locations of the craters in (a).

**Table 1. Craters with boulders observed at 35m/px, and the number and the maximum size of the boulders.**

| Crater Name | Lat. | Lon. (ºE) | D(km) | $n_{d>105m}$ | $n_{d>140m}$ | Maximum boulder diameter (m) |
|---|---|---|---|---|---|---|
| Abellio | 33.3 | 292.9 | 33.3 | 5 | 5 | 291.21 |
| Achita | 25.8 | 65.8 | 40 | 5 | 5 | 228.2 |
| Annona | -48.2 | 8.5 | 58.4 | 7 | 7 | 298.06 |
| Aristaeus | 23.4 | 97.7 | 36.1 | 7 | 7 | 216.41 |
| Axomama | 22.9 | 132 | 5.1 | 2 | 1 | 143.39 |
| Azacca | -6.8 | 218.2 | 52 | 45 | 34 | 310.74 |
| Besua | -42.4 | 300.2 | 17.7 | 48 | 47 | 310.55 |
| Braciaca | -22.8 | 84.4 | 8.3 | 50 | 34 | 231.98 |
| Cacaguat | -1.2 | 143.6 | 13.9 | 124 | 78 | 258.59 |
| Centeotl | 19 | 141.2 | 5.6 | 5 | 3 | 175 |
| Cozobi | 45.4 | 287.3 | 22.4 | 10 | 10 | 289.79 |
| Dantu | 24 | 137.9 | 127.9 | 10 | 10 | 310.01 |
| Darzamat | -44 | 76.2 | 90.9 | 4 | 4 | 206.74 |
| Doliku | -40.7 | 5.9 | 15.4 | 6 | 6 | 290.88 |
| Duginavi | 39.2 | 3.9 | 152.6 | 2 | 2 | 235.46 |
| Emesh | 11.1 | 158.2 | 20.5 | 47 | 44 | 310.58 |
| Ernutet | 52.8 | 45.1 | 50.7 | 3 | 0 | 130.25 |
| Ezinu | 43 | 195.3 | 116.2 | 3 | 3 | 195.14 |
| Gaue | 30.8 | 86 | 81.6 | 11 | 11 | 288.95 |
| Hamori | -60.8 | 79.1 | 55 | 4 | 4 | 309.24 |
| Haulani | 5.8 | 10.8 | 34.6 | 94 | 73 | 283.11 |
| Ialonus | 48.2 | 168.5 | 16.1 | 30 | 29 | 291.83 |
| Ikapati | 33.8 | 45.5 | 51.2 | 12 | 12 | 309.97 |
| Jacheongbi | -69.2 | 2.3 | 28 | 487 | 458 | 309.67 |
| Juling | -35.9 | 168.4 | 19.2 | 97 | 86 | 268.64 |
| Kait | -4.1 | 0.1 | 203.2 | 5 | 3 | 266.95 |
| Kokopelli | 18.2 | 124.5 | 33.6 | 186 | 185 | 301.29 |
| Kondos | -19.3 | 17.3 | 44.5 | 10 | 10 | 256.91 |
| Kupalo | -39.5 | 173.2 | 27.2 | 81 | 76 | 274.29 |
| Laukumate | 65.1 | 159.3 | 28.9 | 13 | 13 | 251.81 |
| Ninsar | 30.4 | 263.1 | 39.1 | 12 | 12 | 300.68 |
| Nunghui | -54.1 | 272.2 | 21.8 | 366 | 324 | 308.23 |
| Occator | 19.8 | 239.4 | 92.2 | 66 | 63 | 292.69 |
| Oxo | 42.2 | 359.6 | 8.9 | 8 | 5 | 226.62 |
| Rao | 8.1 | 119 | 12.2 | 29 | 28 | 249.35 |
| Ratumaibulu | -67.3 | 77.3 | 18.3 | 109 | 95 | 279.78 |
| Roskva | 58.9 | 333 | 21.4 | 5 | 5 | 257 |
| Sekhet | -66.2 | 255 | 40.6 | 98 | 97 | 294.78 |
| Shennong | 69.1 | 28.2 | 30.1 | 98 | 80 | 310.18 |
| Sintana | -48 | 46.2 | 55.4 | 2 | 1 | 141.07 |
| Tawals | -39.1 | 238 | 8.3 | 23 | 21 | 239.88 |
| Thrud | -71.3 | 31 | 6.7 | 27 | 20 | 252.46 |
| Tupo | -32.2 | 88.3 | 34.6 | 71 | 69 | 292.05 |
| Urvara | -45.4 | 249.2 | 157.8 | 12 | 12 | 302.37 |
| Victa | 36.2 | 301 | 30.3 | 15 | 14 | 291.96 |
| Xevioso | 0.7 | 310.6 | 8.6 | 30 | 28 | 221.28 |
| Unnamed2 | -2.6 | 10.6 | 9 | 42 | 35 | 250.2 |
| Unnamed3 | 34.8 | 37.1 | 6.8 | 10 | 7 | 263.34 |
| Unnamed4 | 58.6 | 58.1 | 10.6 | 16 | 13 | 221.14 |
| Unnamed6 | -10.7 | 86.3 | 5.6 | 13 | 10 | 182.66 |
| Unnamed9 | 50.2 | 270 | 8.9 | 39 | 31 | 246.31 |
| Unnamed11 | -23 | 279 | 16.1 | 336 | 239 | 261.3 |
| Unnamed12 | -35.7 | 306.9 | 7.7 | 23 | 18 | 245.62 |
| Unnamed14 | 18.2 | 344.1 | 9.8 | 57 | 46 | 310.44 |
| Unnamed15 | -67.2 | 244.7 | 9.3 | 34 | 24 | 210.22 |
| Unnamed16 | -34.3 | 266 | 14.3 | 166 | 133 | 257.42 |
| Unnamed17 | -10 | 21 | 19 | 209 | 188 | 309.53 |
| Unnamed18 | 33.9 | 355.7 | 8.6 | 26 | 16 | 296.16 |
| Unnamed19 | -68.2 | 153 | 12.3 | 42 | 40 | 264.19 |
| Unnamed21 | -3.1 | 349.6 | 7.9 | 27 | 26 | 264 |
| Unnamed22 | 17 | 62.4 | 9.1 | 56 | 39 | 278.54 |
| Unnamed23 | -64.4 | 233.2 | 6.1 | 13 | 12 | 253.13 |
| Unnamed24 | -44.7 | 313.9 | 11.5 | 49 | 41 | 294.55 |
| Unnamed26 | 63 | 251.9 | 20.1 | 47 | 43 | 263.66 |
| Unnamed27 | -25.8 | 191.5 | 11.9 | 72 | 48 | 257.6 |
| Unnamed28 | 79.6 | 46.7 | 13.5 | 35 | 24 | 253.81 |
| Unnamed29 | -11.4 | 199.1 | 16 | 37 | 35 | 260.72 |
| Unnamed30 | 50.1 | 238.4 | 21.5 | 102 | 102 | 299.11 |
| Unnamed31 | -5 | 276.5 | 5.2 | 21 | 15 | 207.79 |
| Unnamed32 | 68.7 | 300.6 | 7.2 | 8 | 7 | 211.88 |
| Unnamed33 | -47.8 | 19.6 | 7.6 | 31 | 30 | 267.36 |
| Unnamed34 | -39.3 | 110.8 | 19.7 | 100 | 95 | 309.83 |
| Unnamed36 | 59 | 86.5 | 10.2 | 13 | 11 | 229.54 |
| Unnamed37 | -52.3 | 333.5 | 10.4 | 28 | 26 | 279.06 |
| Unnamed42 | 65.8 | 131.6 | 19.3 | 17 | 17 | 307.59 |
| Unnamed43 | 68.1 | 134.5 | 15.4 | 17 | 17 | 255.01 |
| Unnamed48 | -18 | 162.1 | 9.2 | 4 | 4 | 270.26 |
| Unnamed49 | 33.5 | 299.7 | 5.4 | 9 | 6 | 175.05 |
| Unnamed50 | 2.9 | 309.4 | 10.7 | 6 | 5 | 327.95 |
| Unnamed51 | -24.4 | 278 | 5.1 | 13 | 7 | 198.41 |
| Unnamed52 | -53.7 | 32.8 | 3.8 | 4 | 1 | 197.51 |
| Unnamed53 | 4 | 151.6 | 6.8 | 11 | 11 | 301.26 |
| Unnamed54 | 24.1 | 20 | 19 | 1 | 1 | 257.18 |
| Unnamed55 | -35.1 | 292.3 | 6.1 | 38 | 31 | 248.61 |
| Unnamed56 | 20.5 | 7.7 | 9.6 | 8 | 8 | 246.67 |
| Unnamed57 | -57.3 | 264 | 10.2 | 3 | 3 | 243.89 |
| Unnamed58 | -60.9 | 260.8 | 7.4 | 4 | 2 | 162.18 |

**Table 2. Craters with boulders observed at 5m/px, and the number and the maximum size of the boulders.**

| Crater Name | Lat. | Lon. (ºE) | D(km) | $n_{d>15m}$ | $n_{d>20m}$ | $n_{d>50m}$ | Maximum boulder diameter (m) |
|---|---|---|---|---|---|---|---|
| Unnamed30 | 50.1 | 238.4 | 21.5 | 639 | 590 | 261 | 216.28 |
| Unnamed59 | 42.8 | 240.3 | 2.3 | 231 | 181 | 0 | 49.39 |
| Unnamed60 | 46.6 | 242.2 | 7.1 | 30 | 20 | 0 | 46.6 |
| Unnamed61 | 46 | 241.8 | 0.79 | 5 | 3 | 0 | 40.37 |
| Unnamed62 | 45.8 | 242 | 0.52 | 10 | 5 | 0 | 24.78 |
| Unnamed63 | 45.8 | 242.3 | 0.57 | 2 | 0 | 0 | 18.95 |
| Unnamed64 | 46 | 242.9 | 0.66 | 3 | 1 | 0 | 24.11 |
| Unnamed65 | 44.6 | 240.8 | 2.1 | 14 | 14 | 3 | 56.55 |
| Unnamed66 | 40 | 242 | 2.1 | 125 | 117 | 1 | 55.62 |
| Unnamed67 | 37.8 | 242.7 | 0.44 | 10 | 5 | 0 | 34.01 |
| Unnamed68 | 39.1 | 241.1 | 1.5 | 8 | 8 | 0 | 38.78 |
| Unnamed69 | 37.3 | 240.9 | 0.41 | 7 | 4 | 0 | 28.39 |
| Unnamed70 | 37.2 | 240.2 | 0.4 | 4 | 4 | 0 | 24.27 |
| Unnamed71 | 36.9 | 240 | 0.95 | 4 | 3 | 0 | 31.59 |
| Unnamed72 | 37.2 | 242.4 | 0.38 | 9 | 2 | 1 | 62.83 |
| Unnamed73 | 36.7 | 241.8 | 0.46 | 11 | 7 | 0 | 25.5 |
| Unnamed74 | 36.5 | 241.7 | 0.69 | 12 | 10 | 0 | 32.91 |
| Unnamed76 | 35.6 | 244.6 | 0.55 | 9 | 6 | 0 | 24.95 |
| Unnamed77 | 34.8 | 241.6 | 1.7 | 5 | 5 | 0 | 47.28 |
| Unnamed78 | 34.2 | 241.7 | 0.32 | 8 | 0 | 0 | 19.84 |
| Unnamed79 | 33.8 | 245 | 1.78 | 6 | 6 | 0 | 46.17 |
| Unnamed80 | 33.2 | 245.9 | 2.6 | 10 | 10 | 6 | 99.35 |
| Unnamed81 | 33.3 | 241.1 | 0.47 | 15 | 4 | 0 | 21.79 |
| Unnamed82 | 32.3 | 241.7 | 0.31 | 3 | 1 | 0 | 21.29 |
| Unnamed83 | 32.3 | 244.8 | 0.44 | 5 | 2 | 0 | 22.46 |
| Unnamed84 | 32.6 | 238 | 0.28 | 2 | 0 | 0 | 16.53 |
| Unnamed85 | 32.5 | 238.1 | 0.45 | 3 | 1 | 0 | 21.8 |
| Unnamed86 | 32.2 | 238.1 | 0.42 | 5 | 2 | 0 | 26.99 |
| Unnamed87 | 31.5 | 238.2 | 0.31 | 1 | 0 | 0 | 17.67 |
| Unnamed88 | 31.2 | 237.9 | 0.6 | 31 | 13 | 0 | 27.03 |
| Unnamed89 | 34.1 | 243.6 | 1.6 | 3 | 3 | 0 | 36.63 |
| Unnamed90 | 33.9 | 243.7 | 0.43 | 3 | 2 | 0 | 33.67 |
| Unnamed91 | 34.3 | 243.3 | 0.36 | 3 | 3 | 0 | 21.33 |
| Unnamed92 | 29.4 | 234.8 | 1.5 | 3 | 3 | 3 | 93.13 |
| Unnamed94 | 29.9 | 245.1 | 0.34 | 2 | 0 | 0 | 19.89 |
| Unnamed95 | 27.2 | 238.3 | 0.57 | 9 | 7 | 0 | 26.54 |
| Unnamed96 | 27.3 | 238.9 | 0.68 | 3 | 2 | 0 | 21.29 |
| Unnamed97 | 27.5 | 243.8 | 0.63 | 4 | 1 | 0 | 22.85 |
| Unnamed98 | 25.8 | 241.1 | 1.35 | 26 | 20 | 2 | 55.09 |
| Occator | 19.8 | 239.4 | 92.2 | 1836 | 1582 | 304 | 178.04 |
| Unnamed99 | 12.3 | 233.6 | 5.9 | 119 | 119 | 109 | 168.43 |
| Unnamed100 | 12.6 | 246 | 1.9 | 237 | 156 | 0 | 47.74 |
| Unnamed101 | 10.7 | 235.2 | 3 | 4 | 4 | 3 | 109.94 |
| Unnamed102 | 9.4 | 237.6 | 0.61 | 3 | 3 | 0 | 37.55 |
| Unnamed103 | 7.1 | 240.4 | 4.4 | 4 | 4 | 0 | 150.08 |
| Unnamed104 | 8.6 | 240.7 | 0.62 | 24 | 8 | 4 | 29.32 |
| Unnamed105 | 8.3 | 239.1 | 0.51 | 4 | 4 | 0 | 29.02 |
| Unnaemd106 | 6.9 | 246.9 | 0.35 | 7 | 5 | 0 | 27.99 |
| Unnamed107 | 7.3 | 247.9 | 0.26 | 2 | 1 | 0 | 22.66 |
| Unnamed108 | 6.6 | 244.1 | 0.88 | 1 | 1 | 0 | 25.74 |
| Unnamed109 | 5.7 | 237.9 | 4.3 | 11 | 11 | 5 | 58.7 |
| Unnamed110 | 4.2 | 239 | 3.8 | 2 | 2 | 1 | 64.35 |
| Unnamed111 | 4.7 | 249.2 | 1.4 | 15 | 12 | 0 | 30.15 |
| Unnamed112 | 2.3 | 243.4 | 2 | 3 | 3 | 0 | 39.41 |
| Unnamed113 | 4.4 | 236.2 | 2.6 | 2 | 2 | 2 | 65.25 |
| Unnamed114 | 2.1 | 248.4 | 1.7 | 1 | 1 | 0 | 35.27 |
| Unnamed115 | 0.2 | 249.8 | 1.5 | 148 | 104 | 1 | 65.74 |
| Unnamed116 | 0.4 | 249.7 | 0.29 | 5 | 0 | 0 | 18.34 |
| Unnamed117 | 0.9 | 246.7 | 1.2 | 1 | 1 | 0 | 43.95 |
| Unnamed118 | 1.9 | 239.3 | 0.9 | 4 | 4 | 2 | 58.69 |
| Unnamed119 | 2.7 | 237.5 | 0.98 | 14 | 14 | 0 | 43.41 |
| Unnamed120 | 2.5 | 236.9 | 1.8 | 5 | 5 | 0 | 37.21 |
| Unnamed121 | 2 | 237.5 | 1.7 | 23 | 19 | 0 | 37.41 |
| Unnamed122 | 1.1 | 236.5 | 1.7 | 5 | 5 | 0 | 26.98 |
| Unnamed123 | 1.2 | 237.3 | 4.3 | 7 | 6 | 0 | 24.04 |
| Unnamed124 | 1.7 | 237.6 | 2.8 | 2 | 2 | 0 | 26.98 |
| Unnamed125 | 1.5 | 236.9 | 2.3 | 2 | 2 | 0 | 48.9 |
| Unnamed126 | 1.1 | 236.9 | 1.1 | 5 | 3 | 0 | 28.53 |
| Unnamed127 | 0.8 | 237.5 | 1.5 | 13 | 13 | 0 | 34.52 |
| Unnamed128 | 0.7 | 238.3 | 3.6 | 1 | 1 | 1 | 54.82 |
| Unnamed129 | 0.3 | 237.2 | 3 | 28 | 25 | 0 | 43.7 |
| Unnamed130 | 2.2 | 236.3 | 1.6 | 7 | 6 | 0 | 43.73 |
| Unnamed131 | 2.5 | 237.3 | 1.4 | 16 | 16 | 1 | 53.17 |
| Unnamed132 | 1.4 | 232.9 | 1.8 | 3 | 3 | 0 | 48.08 |
| Unnamed133 | 2.6 | 234.1 | 1.4 | 2 | 2 | 2 | 59.51 |
| Unnamed134 | 0 | 226.4 | 1.4 | 5 | 5 | 2 | 63.23 |
| Unnamed135 | 0.5 | 209.7 | 2.3 | 5 | 5 | 5 | 136.64 |
| Unnamed136 | -0.4 | 200.3 | 9.3 | 1 | 1 | 1 | 179.68 |
| Unnamed137 | 3.6 | 229.9 | 5.5 | 2 | 2 | 2 | 148.6 |
| Unnamed138 | 0.1 | 241.6 | 3 | 34 | 25 | 3 | 66.84 |
| Unnamed139 | 0.9 | 244.8 | 0.35 | 6 | 2 | 0 | 25.42 |
| Unnamed140 | 11.3 | 221.8 | 1.2 | 2 | 2 | 2 | 120.31 |
| Unnamed141 | 11.2 | 222.6 | 1.7 | 3 | 3 | 3 | 112.47 |
| Unnamed142 | 0.7 | 233.7 | 3 | 10 | 10 | 8 | 94.24 |
| Unnamed143 | -0.4 | 248.3 | 2.1 | 1 | 1 | 0 | 33.66 |
| Unnamed144 | -0.9 | 248.4 | 2.5 | 28 | 28 | 4 | 58.72 |
| Unnamed145 | -0.8 | 248 | 1.2 | 2 | 2 | 0 | 23.21 |
| Unnamed146 | -0.8 | 246.9 | 1.5 | 6 | 5 | 0 | 34.92 |
| Unnamed147 | -1.1 | 246.2 | 0.7 | 3 | 3 | 0 | 28.63 |
| Unnamed148 | -0.1 | 242.7 | 0.81 | 2 | 2 | 0 | 48.97 |
| Unnamed149 | -1.1 | 242.8 | 0.28 | 1 | 1 | 0 | 21.34 |
| Unnamed150 | -0.7 | 240.5 | 1.4 | 3 | 3 | 0 | 39.82 |
| Unnamed151 | -0.7 | 239.8 | 7 | 2 | 2 | 2 | 63.63 |
| Unnamed152 | -1.2 | 239.7 | 1.4 | 3 | 3 | 2 | 70.1 |
| Unnamed153 | -1.3 | 239.8 | 1.7 | 2 | 2 | 1 | 55.8 |
| Unnamed154 | -1.5 | 240.5 | 1 | 1 | 1 | 0 | 44.58 |
| Unnamed155 | -1.3 | 241 | 0.37 | 2 | 2 | 0 | 34.5 |
| Unnamed156 | -0.2 | 232 | 0.55 | 1 | 1 | 1 | 55.64 |
| Unnamed157 | -0.8 | 223.3 | 1.6 | 2 | 2 | 2 | 85.37 |
| Unnamed158 | -1 | 222 | 0.86 | 1 | 1 | 1 | 71.52 |
| Unnamed159 | -0.3 | 217 | 2.3 | 1 | 1 | 1 | 86.8 |
| Unnamed160 | -2 | 208.1 | 13.5 | 3 | 3 | 5 | 249.98 |
| Unnamed161 | -4.9 | 207.1 | 27.7 | 61 | 61 | 43 | 277.81 |
| Unnamed162 | -2.5 | 213.6 | 3.5 | 2 | 2 | 1 | 61.19 |
| Azacca | -6.8 | 218.2 | 52 | 50 | 50 | 49 | 201.92 |
| Unnamed163 | -6.3 | 214.2 | 1.7 | 3 | 3 | 3 | 130.84 |
| Unnamed164 | -6.5 | 213.4 | 2.8 | 3 | 3 | 3 | 87.69 |
| Unnamed165 | -7 | 213.6 | 1.2 | 1 | 1 | 1 | 98.82 |
| Lociyo | -6.5 | 229.1 | 37.2 | 29 | 29 | 29 | 185.04 |
| Unnamed166 | -3.1 | 234.6 | 2.83 | 4 | 4 | 4 | 197.96 |
| Unnamed167 | -2.1 | 238 | 4.32 | 15 | 15 | 0 | 32.47 |
| Unnamed168 | -2.3 | 239.8 | 1.01 | 2 | 2 | 0 | 48.06 |
| Unnamed169 | -2.1 | 244.7 | 0.35 | 2 | 2 | 0 | 32.47 |
| Unnamed170 | -2.2 | 246.3 | 1.34 | 1 | 1 | 0 | 43.22 |
| Unnamed171 | -3.3 | 246 | 2.05 | 4 | 4 | 0 | 44.6 |
| Unnamed172 | -2.8 | 241.8 | 1.68 | 5 | 5 | 0 | 44.67 |
| Unnamed173 | -3.2 | 242 | 0.45 | 2 | 0 | 0 | 18.69 |
| Unnamed174 | -3.2 | 241.8 | 0.484 | 15 | 5 | 0 | 28.06 |
| Unnamed175 | -3.5 | 241.6 | 0.58 | 3 | 2 | 0 | 38.4 |
| Unnamed176 | -3.6 | 241.9 | 0.349 | 2 | 1 | 0 | 21.36 |
| Unnamed177 | -3.3 | 240.8 | 0.79 | 1 | 1 | 0 | 36.44 |
| Unnamed178 | -3.7 | 237.4 | 1.59 | 7 | 7 | 6 | 96.68 |
| Unnamed179 | -5.4 | 237.7 | 2.16 | 1 | 1 | 1 | 48.25 |
| Unnamed180 | -4.3 | 244 | 0.453 | 5 | 4 | 0 | 35.96 |
| Unnamed181 | -4.5 | 244 | 0.382 | 5 | 4 | 0 | 31.23 |
| Unnamed182 | -4.9 | 245 | 0.281 | 6 | 4 | 0 | 27.59 |
| Unnamed183 | -4.6 | 250.2 | 0.5 | 5 | 1 | 0 | 22.93 |
| Unnamed184 | -5.9 | 249.9 | 0.309 | 3 | 0 | 0 | 19.21 |
| Unnamed185 | -5.8 | 247.4 | 0.88 | 2 | 2 | 0 | 28.09 |
| Unnamed186 | -5.8 | 246.7 | 1.57 | 2 | 2 | 0 | 22.8 |
| Unnamed187 | -6.2 | 243.9 | 1.05 | 1 | 1 | 1 | 50.83 |
| Unnamed188 | -5.7 | 242.7 | 0.98 | 1 | 1 | 0 | 34.04 |
| Unnamed189 | -7.5 | 241.6 | 0.86 | 5 | 5 | 0 | 36.12 |
| Unnamed190 | -7.5 | 242.1 | 0.94 | 39 | 30 | 1 | 68.01 |
| Unnamed191 | -6.8 | 241.5 | 1.27 | 4 | 4 | 1 | 51.05 |
| Unnamed192 | -6.9 | 245.1 | 1.1 | 3 | 3 | 0 | 41.73 |
| Unnamed193 | -7.6 | 246.9 | 1.37 | 2 | 2 | 1 | 52.79 |
| Unnamed194 | -7.1 | 249.9 | 0.483 | 24 | 6 | 0 | 23.7 |
| Unnamed195 | -6.6 | 250.1 | 0.308 | 4 | 3 | 0 | 25.62 |
| Unnamed196 | -6.2 | 250.4 | 0.343 | 1 | 0 | 0 | 18.87 |
| Unnamed197 | -8.5 | 244.6 | 1.1 | 4 | 1 | 0 | 20.31 |
| Unnamed198 | -8.4 | 243.7 | 0.362 | 12 | 4 | 0 | 25.77 |
| Unnamed199 | -7.3 | 207.4 | 2.14 | 6 | 6 | 6 | 180.45 |
| Unnamed200 | -12.4 | 212.7 | 1.57 | 14 | 14 | 12 | 72.08 |
| Unnamed201 | -10.8 | 215 | 2.24 | 3 | 3 | 3 | 100.25 |
| Unnamed202 | -12.1 | 215.3 | 2.37 | 1 | 1 | 1 | 85.61 |
| Unnamed203 | -11.3 | 217.5 | 1.5 | 2 | 2 | 2 | 106.23 |
| Unnamed204 | -11.7 | 217.2 | 20.6 | 2 | 2 | 2 | 124.53 |
| Unnamed205 | -10.7 | 218.3 | 0.57 | 2 | 2 | 1 | 75.2 |
| Unnamed206 | -10.8 | 218.3 | 0.61 | 3 | 3 | 1 | 63.98 |
| Unnamed207 | -9.6 | 218.2 | 3.03 | 18 | 18 | 18 | 118.42 |
| Unnamed208 | -9.3 | 220.2 | 1.72 | 8 | 8 | 6 | 107.13 |
| Unnamed209 | -9.4 | 220.5 | 1.4 | 3 | 3 | 3 | 103.22 |
| Unnamed210 | -10.2 | 219.7 | 22.9 | 20 | 20 | 13 | 69.3 |
| Unnamed211 | -12.5 | 219.4 | 1.26 | 3 | 3 | 0 | 35.83 |
| Unnamed212 | -13 | 219.4 | 6 | 20 | 20 | 8 | 87.11 |
| Unnamed213 | -12.4 | 219.3 | 12.7 | 5 | 5 | 3 | 61.31 |
| Unnamed214 | -10.9 | 227.5 | 3.79 | 1 | 1 | 0 | 41.67 |
| Unnamed215 | -11.2 | 227.8 | 4.75 | 2 | 2 | 0 | 47.12 |
| Unnamed216 | -11.7 | 227.3 | 1.97 | 4 | 4 | 2 | 55.22 |
| Unnamed217 | -11.8 | 226.5 | 8.47 | 8 | 8 | 0 | 46.14 |
| Unnamed218 | -12.7 | 226.8 | 1.58 | 7 | 7 | 0 | 49.71 |
| Unnamed219 | -13 | 226.8 | 17.2 | 40 | 40 | 9 | 87.46 |
| Unnamed220 | -14.4 | 226.8 | 14.6 | 47 | 47 | 8 | 70.91 |
| Unnamed221 | -14 | 228.1 | 2.47 | 4 | 4 | 2 | 56.28 |
| Unnamed222 | -14 | 228.5 | 1.65 | 2 | 2 | 0 | 32.5 |
| Unnamed223 | -14 | 228.7 | 1.57 | 4 | 4 | 2 | 63.03 |
| Unnamed224 | -14.4 | 228.8 | 1.59 | 9 | 9 | 0 | 42.13 |
| Unnamed225 | -14.7 | 228.3 | 0.74 | 2 | 2 | 0 | 43.9 |
| Unnamed226 | -14.6 | 228.8 | 0.372 | 2 | 2 | 0 | 34.27 |
| Unnamed227 | -11 | 236.6 | 2.37 | 4 | 4 | 1 | 54.91 |
| Unnamed228 | -12 | 236.9 | 2.16 | 8 | 8 | 1 | 64.04 |
| Unnamed229 | -12.2 | 237.4 | 0.59 | 5 | 5 | 0 | 36.46 |
| Unnamed230 | -12.6 | 237 | 1.37 | 1 | 1 | 0 | 44.81 |
| Unnamed231 | -12.7 | 237.1 | 0.7 | 5 | 5 | 0 | 48.09 |
| Unnamed232 | -12.6 | 237.4 | 1.57 | 3 | 3 | 0 | 38.81 |
| Unnamed233 | -12.6 | 237.9 | 0.383 | 3 | 2 | 0 | 30.6 |
| Unnamed234 | -11.2 | 241.5 | 1.68 | 4 | 4 | 0 | 32.53 |
| Unnamed235 | -11.4 | 241.6 | 0.7 | 2 | 2 | 0 | 25.44 |
| Unnamed236 | -12 | 241.6 | 1.69 | 16 | 15 | 0 | 37.83 |
| Unnamed237 | -12.2 | 241.9 | 1.53 | 6 | 6 | 1 | 51.4 |
| Unnamed238 | -11.4 | 243.1 | 1.3 | 2 | 2 | 0 | 23.44 |
| Unnamed239 | -11.4 | 244.1 | 0.86 | 12 | 6 | 0 | 26.65 |
| Unnamed240 | -10.6 | 249.1 | 0.459 | 1 | 1 | 0 | 21.86 |
| Unnamed241 | -10.8 | 249.4 | 1.47 | 2 | 1 | 0 | 34.85 |
| Unnamed242 | -10.5 | 251 | 0.441 | 3 | 1 | 0 | 23.99 |
| Unnamed243 | -13.9 | 248.8 | 0.58 | 12 | 2 | 0 | 27.1 |
| Unnamed244 | -13.5 | 247.3 | 2.2 | 2 | 2 | 1 | 64.7 |
| Unnamed245 | -13.7 | 245.4 | 1.14 | 4 | 4 | 0 | 31.12 |
| Unnamed246 | -13.9 | 245.3 | 1.03 | 5 | 4 | 0 | 35.11 |
| Unnamed247 | -13.7 | 245 | 1.07 | 5 | 4 | 0 | 30.75 |
| Unnamed248 | -13.7 | 244.8 | 0.95 | 5 | 5 | 0 | 32.62 |
| Unnamed249 | -14 | 244.9 | 0.93 | 2 | 2 | 0 | 35.33 |
| Unnamed250 | -12.9 | 244.4 | 0.449 | 8 | 2 | 0 | 28.22 |
| Unnamed251 | -14.7 | 245 | 1.38 | 11 | 9 | 0 | 42.1 |
| Unnamed252 | -13.4 | 243.8 | 1.33 | 3 | 3 | 0 | 36.88 |
| Unnamed253 | -13.2 | 241.9 | 0.62 | 3 | 3 | 0 | 31.02 |
| Unnamed254 | -15 | 242.3 | 0.87 | 7 | 6 | 0 | 26.14 |
| Unnamed255 | -14.8 | 238.5 | 3.23 | 15 | 14 | 0 | 46.26 |
| Unnamed256 | -14.9 | 230.4 | 1.81 | 2 | 2 | 0 | 42.56 |
| Unnamed257 | -13.7 | 218.7 | 6.29 | 6 | 6 | 3 | 81.24 |
| Unnamed258 | -14.6 | 218.9 | 11.2 | 22 | 22 | 5 | 69.92 |
| Unnamed259 | -13.9 | 217.5 | 14.3 | 15 | 15 | 7 | 79.87 |
| Unnamed260 | -13.2 | 218.8 | 20.2 | 4 | 4 | 1 | 63.67 |
| Unnamed261 | -14.8 | 215.2 | 2.04 | 4 | 4 | 2 | 57.56 |
| Unnamed262 | -16.3 | 215.9 | 7.21 | 8 | 8 | 6 | 74.15 |
| Unnamed263 | -15.5 | 221.9 | 4.06 | 3 | 3 | 2 | 61.05 |
| Unnamed264 | -17.4 | 229.5 | 12.4 | 43 | 43 | 0 | 44.35 |
| Unnamed265 | -15.5 | 231.2 | 0.61 | 3 | 3 | 0 | 47.26 |
| Unnamed266 | -16.6 | 231.3 | 1.76 | 5 | 5 | 0 | 34 |
| Unnamed267 | -16 | 231.6 | 1.33 | 2 | 2 | 2 | 50.85 |
| Unnamed268 | -16.3 | 231.3 | 6.86 | 9 | 9 | 0 | 39.72 |
| Unnamed269 | -16.8 | 237.7 | 2.12 | 13 | 13 | 1 | 55.75 |
| Unnamed270 | -17.1 | 241.8 | 0.87 | 4 | 4 | 1 | 51.78 |
| Unnamed271 | -16.6 | 242.6 | 1.8 | 4 | 4 | 0 | 30.88 |
| Unnamed272 | -16 | 242.9 | 1.04 | 3 | 1 | 0 | 29.99 |
| Unnamed273 | -15.7 | 248.6 | 0.56 | 22 | 6 | 0 | 21.72 |
| Unnamed274 | -15.5 | 248.8 | 2.31 | 3 | 3 | 0 | 30.69 |
| Unnamed275 | -18.6 | 251.7 | 0.94 | 184 | 69 | 0 | 37.59 |
| Unnamed276 | -19.8 | 249.7 | 1.4 | 45 | 33 | 0 | 49.28 |
| Unnamed277 | -19.7 | 237.2 | 2.94 | 34 | 30 | 0 | 42.73 |
| Unnamed278 | -17.6 | 240 | 1.01 | 7 | 6 | 0 | 33.53 |
| Unnamed279 | -17.9 | 232 | 1.36 | 3 | 3 | 0 | 44.62 |
| Unnamed280 | -18.6 | 229.8 | 0.175 | 8 | 8 | 0 | 45.82 |
| Unnamed281 | -21.6 | 217.4 | 2.18 | 2 | 2 | 1 | 85.61 |
| Unnamed282 | -21 | 219.7 | 0.83 | 1 | 1 | 0 | 46.16 |
| Unnamed283 | -20.6 | 219.8 | 2.27 | 1 | 1 | 0 | 38.37 |
| Unnamed284 | -20.2 | 220.1 | 0.68 | 2 | 2 | 0 | 38.73 |
| Unnamed285 | -20.5 | 220.2 | 0.64 | 2 | 2 | 0 | 34.7 |
| Unnamed286 | -20.5 | 220.2 | 0.311 | 3 | 3 | 0 | 31.39 |
| Unnamed287 | -21.2 | 232.4 | 0.392 | 4 | 4 | 0 | 35.32 |
| Unnamed288 | -20.8 | 233.3 | 0.81 | 1 | 1 | 0 | 45.84 |
| Unnamed289 | -21 | 232.9 | 1.73 | 2 | 2 | 1 | 52.33 |
| Unnamed290 | -20.7 | 233.1 | 0.345 | 1 | 1 | 0 | 26.34 |
| Unnamed291 | -20.9 | 233.1 | 9.11 | 8 | 8 | 0 | 38.16 |
| Unnamed292 | -21 | 239.1 | 1.7 | 2 | 2 | 0 | 35.34 |
| Unnamed293 | -21.1 | 242.8 | 1.6 | 269 | 199 | 0 | 43.49 |
| Unnamed294 | -21.1 | 243.7 | 1.24 | 1 | 0 | 0 | 15.9 |
| Unnamed295 | -20.8 | 244.8 | 0.67 | 30 | 10 | 0 | 24.16 |
| Unnamed296 | -20.7 | 244.7 | 0.387 | 1 | 0 | 0 | 15.49 |
| Unnamed297 | -20.5 | 249.7 | 0.88 | 26 | 14 | 0 | 29.69 |
| Unnamed298 | -22 | 242.9 | 1.27 | 1 | 1 | 0 | 30.48 |
| Unnamed299 | -22.1 | 221.7 | 1.12 | 3 | 3 | 2 | 53.96 |
| Unnamed300 | -23.2 | 222.7 | 1.17 | 3 | 3 | 0 | 48.43 |
| Unnamed301 | -24.7 | 241 | 0.66 | 12 | 12 | 0 | 27.92 |
| Unnamed302 | -26.2 | 233.9 | 1.39 | 14 | 14 | 0 | 42.56 |
| Unnamed303 | -25.9 | 221.6 | 1.09 | 18 | 18 | 1 | 56.67 |
| Unnamed304 | -26.5 | 221.2 | 2.88 | 6 | 6 | 1 | 58.7 |
| Unnamed305 | -27 | 230.4 | 0.392 | 2 | 2 | 0 | 37.47 |
| Unnamed306 | -26.6 | 230.3 | 2.19 | 2 | 2 | 0 | 42.66 |
| Unnamed307 | -28.4 | 247.4 | 0.52 | 1 | 0 | 0 | 19.39 |
| Unnamed308 | -29.8 | 245.5 | 1.23 | 42 | 38 | 0 | 36.36 |
| Unnamed309 | -28.9 | 246.5 | 0.56 | 2 | 2 | 0 | 18.16 |
| Unnamed310 | -29.9 | 246.8 | 0.9 | 2 | 1 | 0 | 27.67 |
| Unnamed311 | -34.9 | 253.9 | 2.23 | 5 | 5 | 5 | 78.64 |
| Unnamed312 | -38.6 | 245.9 | 10.5 | 5 | 5 | 5 | 76.22 |
| Unnamed313 | -40.4 | 246.1 | 0.64 | 3 | 3 | 0 | 37.68 |
| Unnamed314 | -41.4 | 247.5 | 0.98 | 2 | 2 | 0 | 41.69 |
| Unnamed315 | -42.2 | 245.9 | 1.11 | 8 | 8 | 0 | 48.76 |
| Unnamed316 | -45.2 | 255.1 | 2.45 | 2 | 2 | 2 | 91.02 |
| Unnamed317 | -42.9 | 246.2 | 0.99 | 12 | 12 | 1 | 56.59 |
| Unnamed318 | -47 | 248.3 | 4.44 | 23 | 23 | 20 | 117.3 |
| Unnamed319 | -46 | 239.2 | 1.62 | 3 | 3 | 2 | 68.8 |
| Unnamed320 | -46.4 | 240.7 | 0.96 | 1 | 1 | 0 | 48.96 |
| Unnamed321 | -46.9 | 241.7 | 0.464 | 2 | 2 | 1 | 50.47 |
| Unnamed322 | -48.2 | 254.9 | 1.73 | 28 | 28 | 19 | 98.43 |
| Unnamed323 | -49.8 | 254.8 | 1 | 1 | 1 | 1 | 54.23 |
| Unnamed324 | -52.8 | 250.5 | 2.29 | 4 | 4 | 4 | 118.46 |
| Unnamed325 | -56.5 | 226.5 | 2.97 | 57 | 57 | 11 | 90.86 |
| Urvara | -45.4 | 249.2 | 157.8 | 374 | 340 | 175 | 198.91 |

**Table 3. Crater ages and the presence of boulders**

| Name | LDM(Ma) | ADM(Ma) | 35m/px*1 | 5m/px*2 | Reference |
|---|---|---|---|---|---|
| **Achita** | 570±60 | 160±20 | yes | — | Pasckert et al. (2018) |
| **Azacca** | 75.9±15.0 | 45.8±5.0 | yes | yes | Schmedemann et al. (2016) |
| **Begbalel** | | 920±50 /2000(±600, -500) | no | — | Toyokawa et al. (2022) |
| **Cacaguat** | 3.29±0.59 | 3.29±0.59 | yes | — | Williams et al. (2018) |
| **Chaminuka** | 990±60 | | no | — | Toyokawa et al. (2022) |
| **Coniraya** | 1500±90 | | no | — | Toyokawa et al. (2022) |
| **Dantu** | 77.7±8.0 | 22.9±2.3 | yes | — | Williams et al. (2018) |
| **Darzamat** | 1300±90 | | yes | — | Toyokawa et al. (2022) |
| **Duginavi** | 1200±60 | | yes | — | Toyokawa et al. (2022) |
| **Ernutet** | 1600±200 | 420±60 | yes | — | Pasckert et al. (2018) |
| **Ezinu** | 990±80 | | yes | — | Toyokawa et al. (2022) |
| **Gaue** | 162.0±21.0 | 64.5±5.2 | yes | — | Schmedemann et al. (2016) |
| | 220±20 | 69±7 | | | Pasckert et al. (2018) |
| **Haulani** | 1.5-3.5 | 2.4-7.4 | yes | — | Schmedemann et al. (2016) |
| | 1.96±0.17 | 1.67±0.28 | | | Krohn et al. (2018) |
| **Ikapati** | 17.0-76.4 | 17.5-37.4 | yes | — | Schmedemann et al. (2016) |
| | 36±5 | 20±5 | | | Paskert et al. (2018) |
| **Kerwan** | 1500±40 | | no | — | Toyokawa et al. (2022) |
| **Kirnis** | 1500±80 | | no | — | Toyokawa et al. (2022) |
| **Kumitoga** | 1200±50 | | no | — | Toyokawa et al. (2022) |
| **Liber** | 440±0.04 | 180±20 | no | — | Pasckert et al. (2018) |
| **Meanderi** | 570±40 | | no | — | Toyokawa et al. (2022) |
| **Mondamin** | 640±50 | | no | — | Toyokawa et al. (2022) |
| **Ninsar** | 136±9.2 | 87±5.9 | yes | — | Scully et al. (2018) |
| **Occator** | 41.8±7 | 12.7±2.7 | yes | yes | Neesemann et al. (2019) |
| **Oxo** | 0.5±0.2 | 0.5±0.2 | yes | — | Schmedemann et al. (2016) |
| | 3±0.9 | 4.3±1 | | | Hughson et al. (2018) |
| **Rao** | 33.3±2.5 | 30.4±2.3 | yes | — | Williams et al. (2018) |
| **Shennong** | 12±1 | 10±1 | yes | — | Ruesch et al. (2018) |
| **Tupo** | 48.3±6.5 | 35.9±3.0 | yes | — | Schmedemann et al. (2016) |
| | 29±2 | 24±2 | | | Schulzeck et al. (2018) |
| **Unnamed4** | 12±2 | 15±2 | yes | — | Paskert et al. (2018) |
| **Unnamed17** | 3.0±0.7 | 3.0±0.4 | yes | — | Schmedemann et al. (2016) |
| **Unnamed26** | 10.9±1.3 | 7.04±1.4 | yes | — | Scully et al. (2018) |
| **Unnamed28** | 26±2 | 20±5 | yes | — | Ruesch et al. (2018) |
| **Unnamed30** | 68.6±2.5 | 49.6±4.1 | yes | yes | Scully et al. (2018) |
| **Unnamed36** | 53±4 | 55±4 | yes | — | Paskert et al. (2018) |
| **Unnamed44** | 99±3 | 39±7 | no | — | Ruesch et al. (2018) |
| **Urvara** | 242.0±16.0 | 110.0±7.2 | yes | yes | Schmedemann et al. (2016) |
| **Vinotonus** | 2000±200 | | no | — | Toyokawa et al. (2022) |
| **Yalode** | 570±20 | | no | — | Toyokawa et al. (2022) |
| **Zadeni** | 930±70 | | no | — | Toyokawa et al. (2022) |

***1 The presence and absence of boulders around each crater are indicated with 'yes' or 'no', respectively.**

***2 Craters that were not observed by the 5m/px images are indicated with a hyphen.**

**Table 4. Craters without the 5m/px-resolved boulders.**

| Crater Name | Lat. | Lon. (°E) | D(km) |
|---|---|---|---|
| **Nepen** | 6.28 | 220.6 | 27.4 |
| **Unnamed326** | 54.3 | 236.7 | 11.5 |
| **Unnamed327** | 37.0 | 235.2 | 17.5 |
| **Unnamed328** | 43.2 | 242.2 | 10.2 |
| **Unnamed329** | -14.9 | 220.7 | 15.6 |
| **Unnamed330** | 1.27 | 204.8 | 49.3 |
| **Unnamed331** | -0.194 | 214.1 | 27.2 |
| **Unnamed332** | -13.4 | 215.6 | 18.8 |
| **Unnamed333** | -31.2 | 246.5 | 15.4 |

## Reference

Arakawa, M., M. Okazaki, M. Nakamura, M. Jutzi, M. Yasui, S. Hasegawa, (2022) Dispersion and shattering strength of rocky and frozen planetesimals studied by laboratory experiments and numerical simulations. Icarus 373, 114777.

Ballouz, R.-L. et al. (2020) Bennu's near-Earth lifetime of 1.75 million years inferred from craters on its boulders. Nature 587, 205-209.

Basilevsky, A.T., J.W. Head, F. Hörz (2013) Survival times of meter-sized boulders on the surface of the Moon. Planetary and Space Science 89, 118-126.

Basilevsky, A.T., J.W. Head, F. Hörz, K. Ramsley (2015) Survival times of meter-sized rock boulders on the surface of airless bodies. Planetary and Space Science 117, 312-328.

Bart, G.D., H.J. Melosh (2007) Using lunar boulders to distinguish primary from distant secondary impact craters. GRL 34, Issue 7.

Bart, G.D., H.J. Melosh (2010) Distribution of boulders ejected from lunar craters. Icarus 209, pp. 337-357.

Binder, A.B., R.E. Arvidson, E.A. Guinness, K.L. Jones, T.A. Mutch, E.C. Morris, D.C. Pieri, C. Sagan (1977) The geology of the Viking Lander 1 site. JGR 82, pp. 4439-4451.

Buczkowski, D.L. et al. (2016) The geomorphology of Ceres, Science 353, Issue 6303, aaf4332

Chapman, C.R., W.J. Merline, P.C. Thomas, J. Joseph, A.F. Cheng, N. Izenberg (2002) Impact history of Eros: Craters and boulders. Icarus, 155 (1), pp. 104-118.

Christensen, P.R. et al. (2009) JMARS – a planetary GIS American Geophysical Union, Fall Meeting 2009, Abstract #IN22A-06.

Cintala, M. J., J.B. Garvin, S.J. Wetzel (1982) The distribution of blocks around a fresh lunar mare crater. Proc. Lunar Planet. Sci. Conf. 13, 100–101.

Cintala, M.J., K.M. McBride (1995) Block distributions on the lunar surface: A comparison between measurements obtained from surface and orbital photography. NASA Technical Memorandum 104804.

Clauset, A., C.R. Shalizi, M.E.J. Newman (2009) Power-Law Distributions in Empirical Data. SIAM Review 51, Issue 4, 10.1137/070710111.

Daly, R.T. et al. (2023) Successful kinetic impact into an asteroid for planetary defense. Nature 616, 443–447.

Delbo, M. et al. (2014) Thermal fatigue as the origin of regolith on small asteroids. Nature 508, 233-236.

Dellagiustina, D.N. et al. (2019) Properties of rubble-pile asteroid (101955) Bennu from OSIRIS-REx imaging and thermal analysis. Nature Astronomy 3, 341-351.

De Sanctis, M.C. et al. (2015) Ammoniated phyllosilicates with a likely outer Solar System origin on (1) Ceres. Nature 528, 241-244.

De Sanctis, M.C. et al. (2018) Ceres's global and localized mineralogical composition determined by Dawn's Visible and Infrared Spectrometer (VIR). Meteoritics & Planetary Science 53, Issue 9, 1844-1865.

Dombard, A.J., O.S. Barnouin, L.M. Prockter, P.C. Thomas (2010) Boulders and ponds on the Asteroid 433 Eros. Icarus 210 (2), pp. 713-721.

Formisano, M., M.C. De Sanctis, G. Magni, C. Federico, M.T. Capria (2016) Ceres water regime: surface temperature, water sublimation and transient exo(atmo)sphere. Monthly Notices of the Royal Astronomical Society 455, Issue 2, 1892-1904

Fujiwara, A., P. Cerroni, E. Ryan, M. Di. Martino, K. Holsapple, K. Housen, (1989) Experiments and Scaling Laws for Catastrophic Collisions, University of Arizona Press, Asteroids II, p. 240-265.

Fujiwara, A. et al. (2006) The Rubble-Pile Asteroid Itokawa as Observed by Hayabusa, Science 312, Issue 5778, pp. 1330-1334

Ghent, R.R., P.O. Hayne, J.L. Bandfield, B.A. Campbell, C.C. Allen, L.M. Carter, D.A. Paige (2014) Constraints on the recent rate of lunar ejecta breakdown and implications for crater ages. Geology 42, pp. 1059-1062.

Grindrod, P.M., J.M. Davis, S.J. Conway (2021) Active Boulder Falls in Terra Sirenum, Mars: Constraints on Timing and Causes. GRL 48, Issue 20 / e2021GL094817.

Harris, A.W. (1996) The rotation rates of very small asteroids: Evidence for ‘rubble pile’ structure. Lunar Planet. Sci. XXVII, p. 493. Lunar and Planetary Institute, Houston.

Hartmann, W.K. (1969) Terrestrial, lunar, and interplanetary rock fragmentation. Icarus 10, Issue 2, 201-213.

Hiesinger, H. et al. (2016) Cratering on Ceres: Implications for its crust and evolution, Science 353, Issue 6303, aaf4759

Hirata, N., R. Morishima, K. Ohtsuki, A.M. Nakamura, (2022) Disruption of Saturn’s ring particles by thermal stress. Icarus 378, 114919.

Hörz, F., E. Schneider, D.E. Gault, J.B. Hartung, D.E. Brownlee (1975a) Catastrophic rupture of lunar rocks: a Monte-Carlo simulation. The Moon 13, pp. 235-238.

Hörz, F., D.E. Brownlee, H. Fechtig, J.B. Hartung, D.A. Morrison, G. Neukum, E. Schneider, J.F. Vedder, D.E. Gault (1975b) Lunar microcraters—Implications for the micrometeoroid complex. Planetary and Space Science 23, pp. 151-172.

Hörz, F., M. Cintala (1997) Impact experiments related to the evolution of planetary regoliths, Meteoritics & Planetary Science 32, pp. 179-209.

Housen, K.R., K.A. Holsapple (1999) Scale effects in strength-dominated collisions of rocky asteroids. Icarus 142, pp. 21-33.

Hughson, K.H.G. et al. (2018) The Ac-5 (Fejokoo) quadrangle of Ceres: Geologic map and geomorphological evidence for ground ice mediated surface processes. Icarus 316, 63-83

Ikeda, A., H. Kumagai, T. Morota (2022) Topographic Degradation Processes of Lunar Crater Walls Inferred From Boulder Falls. JGR Planets 127, Issue 10 / e2021JE007176.

Jiang, Y., J. Ji, J. Huang, S. Marchi, Y. Li, W.-H. Ip (2015) Boulders on asteroid Toutatis as observed by Chang'e-2. Scientific Reports 5, 16029.

Krishna, N., P.S. Kumar (2016) Impact spallation processes on the Moon: A case study from the size and shape analysis of ejecta boulders and secondary craters of Censorinus crater. Icarus 264, 274-299.

Krohn, K. et al. (2018) The unique geomorphology and structural geology of the Haulani crater of dwarf planet Ceres as revealed by geological mapping of equatorial quadrangle Ac-6 Haulani. Icarus 316, 84-98.

Kuiper, G.P. (1965) The surface structure of the Moon. The Nature of the Lunar Surface, The Johns Hopkins Press, Baltimore, pp. 99-105.

Küppers, M. et al. (2012) Boulders on Lutetia. Planetary and Space Science 66, Issue 1, 71-78.

Kumar, P.S. et al. (2016) Recent shallow moonquake and impact-triggered boulder falls on the Moon: New insights from the Schrödinger basin, JGR Planets 121, Issue 2, pp. 147-179

Lee, S., P. Thomas, J. Veverka (1986) Phobos, Deimos, and the Moon: Size and distribution of crater ejecta blocks. Icarus, 68 (1), pp. 77-86.

Lee, P. et al. (1996) Ejecta Blocks on 243 Ida and on Other Asteroids. Icarus 120, Issue 1, 87-105.

Li, Y., Y. Zhao (2023) The Toutatis (4179) Boulders: Shallow in Size Distribution and Shape Statistics. Solar System Research 57, pp. 495-504.

Martens, H.R., A.P. Ingersoll, S.P. Ewald, P. Helfenstein, B. Giese (2015) Spatial distribution of ice blocks on Enceladus and implications for their origin and emplacement. Icarus 245, 162-176.

McCord, T.B., J.C. Castillo-Rogez (2018) Ceres's internal evolution: The view after Dawn. Meteoritics & Planetary Science 53, Issue 9 1778-1792.

McEwen, A.S. et al., (2007) Mars Reconnaissance Orbiter's High Resolution Imaging Science Experiment (HiRISE), JGR Planets 112, E05S02.

Melosh, H.J. (1984) Impact ejection, spallation, and the origin of meteorites, Icarus 59, Issue 2, pp. 234-260

Melosh, H.J. (1989) Imapct cratering: a geologic process, Oxford University Press, Oxford: Clarendon Press

Michel, P. et al. (2020) Collisional formation of top-shaped asteroids and implications for the origins of Ryugu and Bennu. Nature Communications 11, 2655.

Michikami, T. et al. (2008) Size-frequency statistics of boulders on global surface of asteroid 25143 Itokawa. Earth, Planets and Space 60, 13-20.

Michikami, T. et al. (2019) Boulder size and shape distributions on asteroid Ryugu. Icarus 331, 179-191.

Michikami, T., A. Hagermann (2021) Boulder sizes and shapes on asteroids: A comparative study of Eros, Itokawa and Ryugu. Icarus 357, 114282.

Miyamoto, H. et al. (2007) Regolith Migration and Sorting on Asteroid Itokawa. Science 316, Issue 5827, 1011-1014.

Molaro, J.L., C.P. McKay (2010) Processes controlling rapid temperature variations on rock surfaces. Earth Surface Processes and Landforms 35, Issue 5, 501-507.

Molaro, J., S. Byrne (2012) Rates of temperature change of airless landscapes and implications for thermal stress weathering. JGR Planets 117, E10011.

Molaro, J.L., S. Byrne, J.-L. Le (2017) Thermally induced stresses in boulders on airless body surfaces, and implications for rock breakdown. Icarus 294, 247-261.

Moore, H.J. (1971) Large blocks around lunar craters. Analysis of Apollo 10 Photography and Visual Observations, NASA SP-232, p. 26-27.

Nakamura, A.M. et al. (2008) Impact process of boulders on the surface of asteroid 25143 Itokawa—fragments from collisional disruption. Earth, Planets and Space 60, 7-12.

Nathues, A., H. Sierks, P. Gutierrez-Marques, J. Ripken, I. Hall, I. Buettner, M. Schaefer, U. Chistensen (2016) Dawn FC2 Calibrated Ceres images V1.0, DAWN-A-FC2-3-RDR-CERES-IMAGES-V1.0, NASA Planetary Data System.

Neesemann, A. et al. (2019) The various ages of Occator crater, Ceres: Results of a comprehensive synthesis approach. Icarus 320, 60-82.

Newman, M.E.J. (2005) Power laws, Pareto distributions and Zipf's law. Contemporary Physics 46, Issue 5 323-351.

Pajola, M. et al. (2015) Size-frequency distribution of boulders ≥7 m on comet 67P/Churyumov-Gerasimenko. A&A 583, A37.

Pajola, M. et al. (2016) Size-frequency distribution of boulders ≥10 m on comet 103P/Hartley 2. A&A 585, A85.

Pajola, M., A. Luccetti, L. Senter, G. Cremonese (2021) Blocks Size Frequency Distribution in the Enceladus Tiger Stripes Area: Implications on Their Formative Processes. Universe 2021, 7(4), 82.

Park, R.S. et al. (2016) A partially differentiated interior for (1) Ceres deduced from its gravity field and shape. Nature 537, 515-517.

Pasckert, J.H. et al. (2018) Geologic mapping of the Ac-2 Coniraya quadrangle of Ceres from NASA's Dawn mission: Implications for a heterogeneously composed crust. Icarus 316, 28-45.

Pravec, P., A.W. Harris (2000) Fast and Slow Rotation of Asteroids. Icarus 148(1):12–20.

Prettyman, T.H. et al. (2016) Extensive water ice within Ceres' aqueously altered regolith: Evidence from nuclear spectroscopy. Science 355, Issue 6320, 55-59.

Roatsch, Th. et al. (2017) High-resolution Ceres Low Altitude Mapping Orbit Atlas derived from Dawn Framing Camera images. Planetary and Space Science 140, 74-79.

Ruesch, O. et al. (2018) Geology of Ceres' North Pole quadrangle with Dawn FC imaging data. Icarus 316, 14-27.

Russell, C.T. et al. (2007) Dawn Mission to Vesta and Ceres. Symbiosis between Terrestrial Observations and Robotic Exploration. Earth, Moon and Planets 101, 65-91.

Russell, C.T., C.A. Raymond (2012) The Dawn Mission to Vesta and Ceres. Space Science Reviews 163, 3-23.

Saito, J., et al. (2006) Detailed Images of Asteroid 25143 Itokawa from Hayabusa, Science 312, Issue 5778, 1341-1344

Scheeres, D.J., D. Britt, B. Carry, K.A. Holsapple (2015) Asteroid Interiors and Morphology. P. Michel, F. DeMeo, W. Bottke (eds) Asteroids IV, University of Arizona Press, Tucson, AZ, pp. 745–766.

Schmedemann, N. et al. (2016) Timing of optical maturation of recently exposed material on Ceres. GRL 43, Issue 23, 11987-11993.

Schröder, S.E., U. Carsenty, E. Hauber, F. Schulzeck, C.A. Raymond, C.T. Russell (2020) The Boulder Population of Asteroid 4 Vesta: Size-Frequency Distribution and Survival Time. Earth and Space Science 8, Issue 2, e2019EA000941.

Schröder, S.E., U. Carsenty, E. Hauber, C.A. Raymond, C.T. Russell (2021) The Brittle Boulders of Dwarf Planet Ceres. The Planetary Science Journal 2, Number 3.

Schulzeck, F. et al. (2018) Geologic mapping of the Ac-11 Sintana quadrangle: Assessing diverse crater morphologies. Icarus 316. 154-166.

Scully, J.E.C. et al. (2018) Ceres' Ezinu quadrangle: a heavily cratered region with evidence for localized subsurface water ice and the context of Occator crater. Icarus 316, 46-62.

Shoemaker, E. M. (1965) Preliminary analysis of the fine structure of the lunar surface in Mare Cognitum, in Ranger VII Part II: Experimenters' Analyses and Interpretations, NASA TR 32-700, pp. 75 – 132, Natl. Aeronaut. and Space Admin., Washington, D. C.

Shoemaker, E.M., E.C. Morris (1970) Physical Characteristics of the Lunar Regolith Determined From Surveyor Television Observations. Radio Science 5, Issue 2, 129-155.

Sullivan, R. et al. (1996) Geology of 243 Ida. Icarus 120, 119–139.

Thomas, P.C. et al. (2000) Phobos: Regolith and ejecta blocks investigated with Mars Orbiter Camera images. JGR 105, Issue E6, 15091-15106.

Thomas, P.C., J. Veverka, M.S. Robinson, S. Murchie (2001) Shoemaker crater as the source of most ejecta blocks on the asteroid 433 Eros. Nature 413, 394-396.

Tomasko, M.G. et al. (2005) Rain, winds and haze during the Huygens probe's descent to Titan's surface. Nature 438, 765–778.

Toyokawa, K., J. Haruyama, N. Hirata, S. Tanaka, T. Iwata (2022) Kilometer-scale crater size-frequency distributions on Ceres. Icarus 377, 114909.

Turcotte, D.L., G. Schubert (2002) Geodynamics Second Edition, Cambridge University Press, pp. 271-276

Uribe-Suárez, D., M. Delbo, P.-O. Bouchard, D. Pino-Muñoz (2021) Diurnal temperature variation as the source of the preferential direction of fractures on asteroids: theoretical model for the case of Bennu. Icarus, 360, 114347.

Veverka, J. et al. (2000) NEAR at Eros: Imaging and spectral results. Science 289, 2088–2097.

Viles, H., B. Ehlmann, C.F. Wilson, T. Cebula, M. Page, M. Bourke (2010) Simulating weathering of basalt on Mars and Earth by thermal cycling. GRL 37, L18201.

Watkins, R.N., B.L. Jolliff, K. Mistick, C. Fogerty, S.J. Lawrence, K.N. Singer, R.R. Ghent (2019) Boulder Distributions Around Young, Small Lunar Impact Craters and Implications for Regolith Production Rates and Landing Site Safety. JGR Planets 124, Issue 11, 2754-2771.

Williams, D.A. et al. (2018) The geology of the Kerwan quadrangle of dwarf planet Ceres: Investigating Ceres' oldest, largest impact basin. Icarus 316, 99-113.